\PassOptionsToPackage{dvipsnames}{xcolor}

\documentclass[acmsmall]{acmart}
\usepackage[utf8]{inputenc}
\usepackage[T1]{fontenc}
\usepackage{microtype}
\usepackage[dvipsnames]{xcolor}
\usepackage[frozencache,cachedir=.]{minted}
\usepackage{caption,subcaption}
\usepackage[export]{adjustbox}
\usepackage{wrapfig}
\usepackage{newunicodechar}
\usepackage{ifthen}
\usepackage{enumitem}
\usepackage{amsmath}
\usepackage{multicol}

\newunicodechar{ε}{$\epsilon$}
\newunicodechar{∩}{$\cap$}
\newunicodechar{∞}{$\infty$}
\newunicodechar{σ}{$\sigma$}
\newunicodechar{∈}{$\in$}
\newunicodechar{∈}{$\in$}
\newunicodechar{∉}{$\notin$}
\definecolor{LightGray}{gray}{0.98}
\newcommand{\para}[1]{\textbf{#1.}}

\AtBeginDocument{%
  \providecommand\BibTeX{{%
    \normalfont B\kern-0.5em{\scshape i\kern-0.25em b}\kern-0.8em\TeX}}}

\setcopyright{cc}
\setcctype{by}
\acmDOI{10.1145/3763146}
\acmYear{2025}
\acmJournal{PACMPL}
\acmVolume{9}
\acmNumber{OOPSLA2}
\acmArticle{368}
\acmMonth{10}
\received{2025-03-21}
\received[accepted]{2025-08-12}

\begin{document}

\title{The Continuous Tensor Abstraction: Where Indices Are Real}

\author{Jaeyeon Won}
\orcid{0000-0002-3082-4348}
\affiliation{%
  \institution{Massachusetts Institute of Technology}
  \city{Cambridge}
  \country{USA}
}
\email{jaeyeon@csail.mit.edu}

\author{Willow Ahrens}
\orcid{0000-0002-4963-0869}
\affiliation{%
  \institution{Georgia Institute of Technology}
  \city{Atlanta}
  \country{USA}
}
\email{ahrens@gatech.edu}

\author{Teodoro Fields Collin}
\orcid{0009-0001-7578-7683}
\affiliation{%
  \institution{Massachusetts Institute of Technology}
  \city{Cambridge}
  \country{USA}
}
\email{Teoc@csail.mit.edu}

\author{Joel S. Emer}
\orcid{0000-0002-3459-5466}
\affiliation{%
  \institution{Massachusetts Institute of Technology}
  \city{Cambridge}
  \country{USA}
}
\affiliation{%
  \institution{NVIDIA}
  \city{Westford}
  \country{USA}
}
\email{emer@csail.mit.edu}

\author{Saman Amarasinghe}
\orcid{0000-0002-7231-7643}
\affiliation{%
  \institution{Massachusetts Institute of Technology}
  \city{Cambridge}
  \country{USA}
}
\email{saman@csail.mit.edu}


\begin{abstract}
This paper introduces the continuous tensor abstraction, allowing indices to take real-number values (e.g., \texttt{A[3.14]}). It also presents continuous tensor algebra expressions, such as $C_{x,y} = A_{x,y} \ast B_{x,y}$, where indices are defined over a continuous domain. This work expands the traditional tensor model to include continuous tensors. Our implementation supports piecewise-constant tensors, on which infinite domains can be processed in finite time. We also introduce a new tensor format for efficient storage and a code generation technique for automatic kernel generation. For the first time, our abstraction expresses domains like computational geometry and computer graphics in the language of tensor programming. Our approach demonstrates competitive or better performance to hand-optimized kernels in leading libraries across diverse applications. Compared to hand-implemented libraries on a CPU, our compiler-based implementation achieves an average speedup of 9.20$\times$ on 2D radius search with $\sim$60$\times$ fewer lines of code (LoC), 1.22$\times$ on genomic interval overlapping queries (with $\sim$18$\times$ LoC saving), and 1.69$\times$ on trilinear interpolation in Neural Radiance Field  (with $\sim$6$\times$ LoC saving).

\end{abstract}



\maketitle

\section{Introduction}
Array programming has been a cornerstone in the history of computing, with its origin dating back to FORTRAN's introduction in 1957~\cite{fortran}. It remains integral to imperative languages like C/C++~\cite{c++}, Java~\cite{java}, Julia~\cite{julia}, and Python~\cite{python}, as well as specialized array-focused languages and frameworks including APL~\cite{apl}, MATLAB~\cite{matlab}, TensorFlow~\cite{tensorflow}, and PyTorch~\cite{pytorch}. The array abstraction is simple, consistent across languages, and widely familiar from introductory programming courses. Many domains rely on multidimensional arrays and share a common set of tools built around it.

Traditional dense tensor programming\footnote{Currently, "tensor programming" is often used interchangeably with "array programming," where "tensor" denotes a multidimensional array. For this paper, we adopt the term "tensor programming."} is guided by the fundamental requirement that all indices are integers. While traditional tensor programming relies on integer indices, it is insufficient for many real-world data, which often involves real number coordinates—common in domains such as 3D deep learning~\cite{pointwise,largekernel}, computer graphics~\cite{nerf,plenoxels}, and spatial databases~\cite{shapely,geos}. \color{black}
Unfortunately, without a common abstraction, support for such data has fragmented across domains, with each domain relying on its own tools. We believe that continuous domain programs can be unified under the umbrella of tensor programming. A simple, uniform abstraction would lower the learning curve for programmers and amortize the development cost across a wide range of applications.
\color{black}

\textbf{In this work, we challenge three assumptions that restrict tensor programming to integer coordinates and show it can support real coordinates.} While recent works have implicitly questioned the first two assumptions, our work challenges all three. To our knowledge, we are the first to explicitly address the third assumption, leading to our main contribution.

\textbf{Conventional Assumption 1: Computations are performed at all points in the \emph{iteration space}}—the set of all index tuples within the tensor's shape. This includes both \textit{effectual computations} (those contributing to the final output) and \textit{ineffectual computations} (those that do not affect the output, e.g., \verb|out+=in*0|). Like other sparse compilers~\cite{taco,mlirdialect,finch}, we break this assumption by computing only at effectual points and skipping ineffectual points without sacrificing correctness.

\textbf{Conventional Assumption 2: The iteration space is finite} because every point in the iteration space must project into coordinates in the tensors. Traditionally, computations at out-of-bounds indices were considered invalid or caused errors. By conceptualizing tensors as implicitly holding zeros at out-of-bound indices, we can extend the iteration space—even to infinity—simplifying programming models and avoiding out-of-bounds errors. This is true because computations at these indices are ineffectual and can be skipped. Some prior tensor compilers, such as Halide~\cite{halide} and Finch~\cite{finch}, implicitly break this assumption. This work also relies on breaking this assumption. 

\textbf{Conventional Assumption 3: \color{black} The iteration space is an integer lattice. \color{black}} Traditionally, the points in an $N$-dimensional iteration space are $N$-tuples of discrete integer coordinates. As an iteration space can be infinite, the domain of the space do not have to be integers either. Thus, we could have real-number coordinates. This allows us to envision the iteration space as continuous rather than discrete, enabling computations at real-valued indices. We generalize computations to operate over real domains, focusing on effectual computations that occur at real-number indices.

In this paper, we extend the tensor programming model by expanding coordinate points from the discrete space of integers to the continuous space of real numbers. Users can write tensor algebra expressions such as $C_x = A_x + B_x$ and $C_{x,y} = A_{x,z} * B_{z,y}$, using real-numbered indices and iterating over a continuous domain. By leveraging a piecewise-constant assumption, we propose implementation methods for storing continuous tensors in memory, performing reduction operations over continuous iteration spaces, and generating efficient code for continuous tensor programs. By expressing applications using the tensor programming model that previously required domain-specific codes, we bring the simplicity of a universal abstraction to those domains.

To our knowledge, this paper is the first to {\bf extend tensor programming to real-numbered indices with continuous iteration space.} This paper includes the following contributions:

\begin{itemize}
  \setlength\itemsep{1pt}
    \item We extend the tensor programming model to iterate over a continuous domain under a piecewise-constant assumption.
    \item We introduce reduction operations specifically designed for the continuous iteration space.
    \item We introduce an efficient code generation mechanism for computations on continuous tensors by extending the fibertree abstraction~\cite{fibertree} and Looplets~\cite{finch, finchoopsla}.
    \item We unify a diverse range of applications across various fields using the continuous tensor abstraction, including bioinformatics, geospatial applications, point cloud processing, and Neural Radiance Fields (NeRF). Writing applications in the continuous tensor abstraction is straightforward and intuitive, requiring $\sim$18$\times$  fewer lines of code in bioinformatics, $\sim$62$\times$  fewer lines of code in geospatial queries, $\sim$101$\times$  fewer lines of code in 3D point cloud convolution, and $\sim$6$\times$ fewer lines of code in trilinear interpolation in NeRF.
    \item Compared to hand-implemented libraries, our compiler-based implementation achieves an average speedup of 9.20$\times$ on radius search queries, 1.22$\times$ on genomic interval overlapping queries, and 1.69$\times$ on trilinear interpolation in NeRF.
\end{itemize}

\section{Motivation}

\begin{figure}[t]
    \begin{subfigure}[t]{0.48\textwidth}
        \centering
        \vspace{-0.1in}
        \includegraphics[width=1\textwidth]{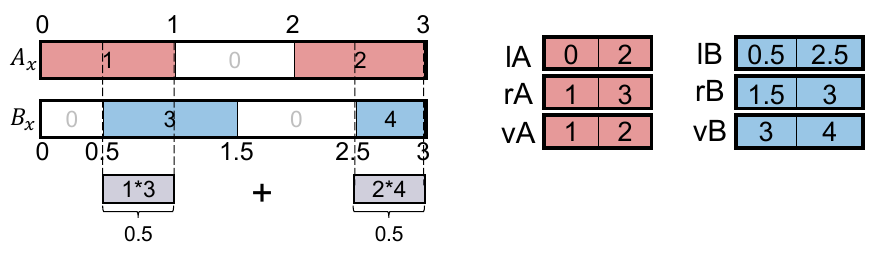}
        \label{fig:pconvvis}
        \vspace{-0.2in}
        \caption{Visualization of interval dot product along with the memory representation of the coordinates (\texttt{lA,rA,lB,rB}) and values (\texttt{vA,vB}).}
        \vspace{0.1in}
    \end{subfigure}\hspace{0.1in}
    \begin{subfigure}[t]{0.4\textwidth}
\begin{minted}[fontsize=\footnotesize,escapeinside=||,mathescape=true,frame=single]{python}
# Format Spec of Continuous Tensors
A = Interval(Element(vA),lA,rA)
B = Interval(Element(vB),rA,rB)
# Continuous Einsum
@einsum val = A[x] * B[x] * d(x,|$\mu_\lambda$|)
\end{minted}
\caption{Interval dot product in continuous tensor abstraction, where $d(x,\mu_\lambda)$ denotes integration over $x$ with Lebesgue measure $\mu_\lambda$.}
        \label{fig:dpb}
    \end{subfigure}

\vspace{0.1in}

    \begin{subfigure}[t]{0.31\textwidth}
\begin{minted}[fontsize=\footnotesize,escapeinside=||,mathescape=true,frame=single]{c++}
field#1(1)[] A, B;
strand dotprod(){
real t = 0;
real tmax = 3;
real step = 0.01;
real val = 0;
update {  
  val += A(t) * B(t) * step;
  t += step;
  if (t > tmax){stabilize;}}} 
\end{minted}
    \caption{Interval dot product \emph{approximately} written in Sci-Vis language Diderot~\cite{diderot} }
        \label{fig:dpc}
    \end{subfigure}
        \hfill
    \begin{subfigure}[t]{0.35\textwidth}
        \vspace{0.0in} 
\begin{minted}[fontsize=\footnotesize,frame=single]{sql}
SELECT
 SUM(
  ST_Length(
   ST_Intersection(A.geom,B.geom)
  ) 
  * A.value * B.value
 )
FROM A
JOIN B
  ON ST_Intersects(A.geom,B.geom)
GROUP BY A.id
\end{minted}
    \caption{Interval dot product written in spatial SQL. }
                       \label{fig:dpd}
    \end{subfigure}
    \hfill
            \begin{subfigure}[t]{0.28\textwidth}
\begin{minted}[fontsize=\footnotesize,escapeinside=||,mathescape=true,frame=single]{python}
pA, pB = 0, 0
while (pA<2 and pB<2):
 lA,rA,vA = A[pA]
 lB,rB,vB = B[pB]
 lAB = max(lA,lB)
 rAB = min(rA,rB)
 if lAB <= rAB:
  sum += (rAB-lAB)*vA*vB
 pA += (rA==rAB) 
 pB += (rB==rAB)
\end{minted}
        \caption{Interval dot product written in array programming with integer indices. }
                \label{fig:dpe}
    \end{subfigure}
    \vspace{-0.1in}
    \caption{\color{black}Illustration of an interval dot product written in four different languages, highlighting how our continuous tensor abstraction provides a concise and straightforward implementation compared to others.\color{black}}
        \label{fig:intervaldp}
\end{figure}


\color{black}

Programming over continuous domains is challenging in current tools. In contrast, our abstraction is simple, intuitive, and concise, while still producing code that matches or outperforms existing approaches. This advantage is especially clear in applications, such as scientific visualization~\cite{diderot,vivaldi} and spatial deep learning~\cite{plenoxels,nerf,pointwise}, which require both (1) geometric operations and (2) numerical computation on geometry‑tied values, a combination rarely supported by current systems.

Figure~\ref{fig:intervaldp} shows the interval dot product, which sums the product of values over the lengths of overlapping intervals, requiring both intersection computation and numerical integration. This operation is useful in many areas, such as bioinformatics. For example, when comparing two genomic intervals $A$ and $B$, it calculates how much of gene $A$’s position overlaps with gene $B$’s position on the chromosome. More operations on genomic intervals are discussed in Section~\ref{subsec:genomic}. Existing models—scientific visualization languages~\cite{diderot,vivaldi}, spatial databases~\cite{geos,shapely}, and tensor frameworks~\cite{numpy,c++}—typically focus only one of two aspects, making such tasks difficult to express.

\textbf{SciVis. } Languages like Diderot~\cite{diderot} support pointwise evaluations on continuous fields, but lack mechanisms for expressing geometric operations like intersections. As shown in Figure~\ref{fig:dpc}, Diderot allows querying field values at specific points on real domain, but cannot compute field intersections directly, requiring users to approximate the interval dot product via Riemann integration.

\textbf{Spatial Database. } Spatial databases~\cite{spatialsql,geos} support geometric queries and efficient spatial structures like R-trees~\cite{strtree}, but are limited in expressing numerical computations. As shown in Figure~\ref{fig:dpd}, spatial SQL easily handles intersections and lengths, but becomes cumbersome for complex computations on multidimensional tensors.

\textbf{Tensor Programming. } Tensor frameworks excel at high-performance numerical computation but assume dense, integer indexing, making them poorly suited for continuous domains. Supporting real-valued coordinates requires custom data structures (e.g., COO formats, interval trees~\cite{format,ncls,ail}) and significant engineering to implement geometric operations like intersection. As shown in Figure~\ref{fig:dpe}, low-level implementations can perform efficient intersection and weighted sums but must be re-written for each new computation.

To bridge this gap, we propose a continuous tensor abstraction that unifies the strengths of existing models. Tensors are defined over real coordinates and have nonzero values on geometrically meaningful regions (e.g., boxes, lines). We generalize tensor indexing and extend Einsums~\cite{einstein_foundation_1916,teaal,numpy} to continuous domains, enabling concise, expressive geometric and numerical computation.

\textbf{Continuous Tensor Abstraction. } As shown in Figure~\ref{fig:dpb}, our abstraction allows users to define data structures using a \emph{level format abstraction}~\cite{format,fibertree} and write computations using an \emph{extended Einsum notation} with real-valued indices. The resulting code is compiled into efficient low-level implementations, equivalent to those shown in Figure~\ref{fig:dpe}. Users simply write a familiar Einsum expression and format specification, yielding a unified programming model for continuous domain that combines both geometric and numerical computation. 


\color{black}

\section{Continuous Tensor Expression Language} \label{sec:lang}

\subsection {Background}
Tensor compilers~\cite{teaal,numpy, halide, taco, tvm,pytorch} provide a syntax similar to Einsums (Einstein summation notation~\cite{einstein_foundation_1916}), a concise way to describe operations like matrix multiply (\(C_{x,y} = \sum_r A_{x,r} * B_{r,y}\)).

An Einsum can be understood as a traversal over all coordinates of the index variables. During this traversal, the specified tensor expression is evaluated at each point and the result is stored in the output. The reduction operator is used when the same output coordinate is referenced more than once. In other words, reductions occur along index variables that do not appear in the output. This typically implies summation over those dimensions, removing the need for explicit summation symbols, as in matrix multiplication (\(C_{x,y} = A_{x,r} * B_{r,y}\)). Reductions, however, are not limited to summation; they can also involve operations like min, max, and other aggregations. 

\subsection{Syntax and Denotational Semantics} \label{subsec:semiring}

\begin{figure}
\resizebox{.9\textwidth}{!}{ 
\begin{minipage}{0.44\textwidth}
    \begin{align*}
        \langle einsum\rangle &::= \langle stmt \rangle \\
        &\quad \mid \langle stmt \rangle * d(\langle rIndex \rangle...; \langle measure\rangle...)
        \\[8pt]
        \langle stmt\rangle &::= \langle tensorName\rangle[\langle index\rangle \dots] = \langle expr\rangle
        \\[8pt]
        \langle expr\rangle &::= \langle val \rangle \\
        &\quad \mid \langle affineIndex\rangle \\
        &\quad \mid \langle tensorName\rangle[\langle affineIndex\rangle \dots] \\
        &\quad \mid (\langle expr\rangle) \\
        &\quad \mid \langle expr\rangle \; \langle op\rangle \; \langle expr\rangle
    \end{align*}
\end{minipage}
\hspace{0.2in}
\begin{minipage}{0.45\textwidth}
    \begin{align*}
        \langle affineIndex\rangle &::= \langle affineTerm\rangle \\
        &\quad \mid \langle affineIndex\rangle + \langle affineTerm\rangle\\
        \langle affineTerm\rangle &::= \langle val\rangle \mid \langle index\rangle \mid \langle val\rangle * \langle index\rangle\\
        \langle tensorName\rangle &::= \langle identifier\rangle \\
        \langle index\rangle &::= \langle identifier\rangle \\
        \langle rIndex\rangle &::= \langle identifier\rangle \\
        \langle op\rangle &::= + \mid - \mid * \mid \mathrm{AND} \mid \mathrm{OR} \dots \\
        \langle measure\rangle &::= \mu_{\land\lor} \mid \mu_{\#} \mid \mu_{\lambda} \mid \dots \\
        \langle val\rangle &::= \langle \text{Real Number}\rangle \\
        \langle identifier\rangle &::= [a{-}zA{-}Z]
    \end{align*}
\end{minipage}
}
\vspace{-0.1in}
\caption{Grammar of our continuous Einsum.}
\label{fig:bnf}
\end{figure}

\begin{figure}
\resizebox{0.83\textwidth}{!}{ 
\begin{minipage}{1.0\textwidth}
\raggedright
\textbf{Einsum}
\begin{align*}
&\llbracket \texttt{tensorName}[\texttt{index}_1,...,\texttt{index}_n] \rrbracket^\text{Env} = ( x \rightarrow \llbracket \texttt{expr} \rrbracket^{\text{Env} \cup \{ \texttt{index}_1,...,\texttt{index}_n \mapsto x \}} 
    : x \in \mathbb{R}^n ) 
\\[8pt]
&\llbracket \texttt{tensorName}[\texttt{index}_1,...,\texttt{index}_n] = \texttt{expr} * d(\texttt{rIndex}_1,...,\texttt{rIndex}_k;  \;\texttt{measure}_1,...,\texttt{measure}_k) \rrbracket^{\text{Env}} = \\
&\quad\quad\quad \text{Let } f(x): \mathbb{R}^{n+k} \to  
    \llbracket \texttt{expr} \rrbracket^{\text{Env} \cup \{ \texttt{index}_1,...,\texttt{index}_n,\texttt{rIndex}_1,...,\texttt{rIndex}_k  \mapsto x \}} \\
    &\quad\quad\quad \text{in Let } g(x): \mathbb{R}^{n} \to  
    \int^{\oplus}_{\texttt{rIndex}_{1...k}} f(\texttt{index}_{1,...,n},\texttt{rIndex}_{1,...,k}) \, d \mu_1(\texttt{rIndex}_1 )... d\mu_k(\texttt{rIndex}_k) \; \text{in } g 
\end{align*}

\begin{align*}
\textbf{Expression}\;\;\;\;\;\;\;\;\;\;\;\;\;\;\;\;\;\;\;\;\;\;\;\;\;\;\;\;\;\;\;\;\;\;\;\;\;\;\;\;\;\;\;\;\;\;\;
\llbracket \texttt{val} \rrbracket^{\text{Env}} &= \texttt{val} \\[8pt]
    \llbracket \texttt{tensorName}[\texttt{affineIndex} \dots] \rrbracket^{\text{Env}} &=
    \text{Env}[\texttt{tensorName}](\llbracket \texttt{affineIndex} \rrbracket^{\text{Env}} \dots) \\[8pt]
    \llbracket (\texttt{expr}) \rrbracket^{\text{Env}} &= 
    \llbracket \texttt{expr} \rrbracket^{\text{Env}}\\[8pt]
    \llbracket \texttt{expr}_1 \texttt{ op } \texttt{expr}_2 \rrbracket^{\text{Env}} &=
    \llbracket \texttt{op} \rrbracket \left( \llbracket \texttt{expr}_1 \rrbracket^{\text{Env}}, \llbracket \texttt{expr}_2 \rrbracket^{\text{Env}} \right)
\end{align*}

\begin{align*}
\textbf{Affine Index} \;\;\;\;\;\;\;\;\;\;\;\;\;\;\;\;\;\;\;\;\;\;\;\;\;\;\;\;\;\;\;\;\;\;\;\;\;\;\;\;\;   \llbracket \texttt{index} \rrbracket^{\text{Env}} &= \text{Env}[\texttt{index}] \\[8pt]
    \llbracket \texttt{val} * \texttt{index} \rrbracket^{\text{Env}} &= 
    \llbracket \texttt{val} \rrbracket^{\text{Env}} * \llbracket \texttt{index} \rrbracket^{\text{Env}} \\[8pt]
    \llbracket \texttt{affineIndex}_1 + \texttt{affineIndex}_2 \rrbracket^{\text{Env}} &= 
    \llbracket \texttt{affineIndex}_1 \rrbracket^{\text{Env}} + \llbracket \texttt{affineIndex}_2 \rrbracket^{\text{Env}}
\end{align*}
\end{minipage}
}
\vspace{-0.1in}
\caption{\color{black}
Denotational semantics of our language. 
We assume an environment \(\mathrm{Env}\) providing:
\textbf{(1)} a mapping from each tensor name to a function \(\mathbb{R}^n \to \mathbb{S}\) (where \(\mathbb{S}\) is a semiring domain), 
\textbf{(2)} a mapping from each index name to a real number, 
and \textbf{(3)} a mapping from each measure name to a corresponding semiring-valued measure \(\mu_i\). 
Each \(\mu_i\) is referenced in the syntax as \(\texttt{measure}_i\) and, when combined with \(\texttt{rIndex}_i\), enables integrating over the specified reduction indices.\color{black}
}

\label{fig:denotational}
\end{figure}

In Figure~\ref{fig:bnf} and \ref{fig:denotational}, we present the syntax and denotational semantics of our language. It closely resembles existing Einsum notation. One notable difference from existing Einsums is that the reduction over continuous domain may sometimes correspond to discrete operations (e.g., summation) and other times to continuous operations (e.g., integration). To handle this, the user must specify a reduction \emph{measure} for each reduction index with \texttt{d(<rIndex>..., <measure>...)}.

Specifically, Einsum statements in our language can take one of the following forms:
\[
C_{x_1, \dots, x_n} = P_{x_1, \dots, x_n} \text{\quad if there is no reduction}
\]
\[
C_{x_1, \dots, x_n} = P_{x_1, \dots, x_n, rx_1, \dots, rx_k} * d(rx_1, \dots, rx_k; \mu_{1}, \dots, \mu_{k}) \text{\quad otherwise}
\]

Here, \( C \) is the output tensor, whose values are computed over the indices \( x_1, \dots, x_n, rx_1, \dots, rx_k \) defined on a real domain. \color{black}The indices \( (rx_1, \dots, rx_k) \in D \) represent the reduction indices (where $D$ is a reduction domain). \color{black} \(P_{x_1,\dots,x_n}\) denotes a general expression that may involve one or more tensors and depends on the indices \( x_1,\dots,x_n \). We adopt subscript notation for consistency with standard tensor indexing; however, this notation does not imply that \(P\) is necessarily a single tensor. In fact, \(P\) may incorporate multiple tensor accesses and semiring operations~\cite{semiring,semiring1,semiring2}. In the denotational semantics presented in Figure~\ref{fig:denotational}, a statement involving reductions is interpreted as follows:
\[
C_{x_1, \dots, x_n} = \int^\oplus_{{rx_1,\dots,rx_k}} P_{x_1, \dots, x_n, rx_1, \dots, rx_k}  d\mu_1(rx_1)  \dots  d\mu_{k}(rx_k)
\]

The operator $\int^\oplus$ represents an integral using semiring addition \( \oplus \), combining the values of \( P \) across the reduction indices. This reduction is weighted by the \emph{semiring-valued measures} \( \mu_1, \dots, \mu_k \) for each reduction index \( rx_1, \dots, rx_k \), respectively.

Calculations are performed within a semiring \( \mathbb{S}=(R,\oplus,\otimes,0,1) \), which is defined by an addition operation \( \oplus \) (associative, commutative, with identity \( 0 \)) and a multiplication operation \( \otimes \) (associative, with identity \( 1 \), distributing over \( \oplus \)). Common examples include the real number semiring (\(\mathbb{R}, +, \times, 0, 1 \)) and the boolean semiring (\(\{T,F\}, \vee , \wedge, F, T \)). To formally define the reduction process, a measure-theoretic approach~\cite{measure1,measure2,measure3} is used. Given a sigma-algebra over the reduction indices \( V \subset \mathbb{P}(D) \) (a power set of $D$), a \emph{semiring-valued measure} \( \mu : V \to \mathbb{S} \) is defined on subsets of \( D \), satisfying \( \mu(\emptyset) = 0 \) and disjoint additivity: \( \mu(A \cup B) = \mu(A) \oplus \mu(B) \) for disjoint subsets \( A \) and \( B \).

Using measures \( \mu_1, \dots, \mu_k \) for each reduction index \(rx_1,\dots,rx_k\), the reduction is rewritten as: 
\[
C_{x_1,\dots,x_n}
\;=\;
\bigoplus_{m \,\in\, \mathbb{S}}
\Bigl[
  (\mu_1 \otimes \dots \otimes \mu_k) (\{(rx_1,\dots,rx_k)\in D : P_{x_1,\dots,x_n,rx_1,\dots,rx_k} = m\})
\Bigr]
\;\otimes\; m.
\]
This expression defines each output \(C_{x_1, \dots, x_n}\) by summing over all possible values \(m \in \mathbb{S}\) that \(P\) can take. Each value \(m\) is weighted by the product of semiring-valued measures associated with the reduction indices. Specifically, since \(m \in \mathbb{S}\), it may represent any element in the semiring: for example, any real number in the real semiring, or \(\text{true}\) and \(\text{false}\) in the boolean semiring. The measure \(\mu\) provides these weights, determining how much each \(m\) contributes to \(C_{x_1, \dots, x_n}\). Finally, the semiring addition \(\oplus\) is used to combine these weighted contributions into a single value.

$\oplus$ and $\otimes$ could represent real addition and multiplication or the logical OR and AND. While this subsection focuses on the real number semiring and the Boolean semiring, the framework can accommodate additional semirings, allowing the language to support various reduction semantics. 

\subsubsection{\textbf{Real Number Semiring $(R,+,*,0,1)$}} For Real number addition (\( \oplus = + \)), users specify a measure \( \mu \) over \( D \), choosing either the \textbf{Lebesgue measure \( \mu_\lambda \)} or the \textbf{counting measure} \( \mu_\# \)~\cite{realanalysis}. The Lebesgue measure, suitable for continuous intervals, is defined as \( \mu_\lambda([a, b]) = b - a \). The counting measure, suitable for discrete points, is defined as follows: for isolated pinpoints \( [a, a] \), \( \mu_\#([a, a]) = 1 \); for intervals like \( [a, b] \) with \( a < b \), \( \mu_\#([a, b]) = \infty \). Users can flexibly model both continuous and discrete accumulations by selecting the proper measure.

Given a function \( f_P\) representing \( P \) evaluated at \( (x_1, \dots, x_n, rx_1, \dots, rx_k) \), we define:
\[
C_{x_1, \dots, x_n} = \int_{rx_1 \dots rx_k} f_P(x_1, \dots, x_n, rx_1, \dots, rx_k) \, d\mu_{1}(rx_1) \dots d\mu_{k}(rx_k)  
\]

\begin{wrapfigure}{r}{.2\textwidth}
    \vspace{-0.25in}
    \includegraphics[width=\linewidth]{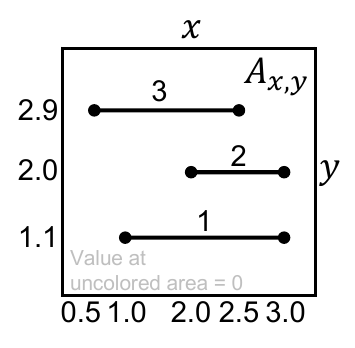}
  \label{fig:2dex}
  \vspace{-0.35in}
\end{wrapfigure} 
In the right example, the Einsum \( s =  A_{x,y} * d(x,y;\mu_{\lambda},\mu_{\#}) \) evaluates $s = \int_y \int_x f_A(x,y) \, d\mu_\lambda(x) \, d\mu_\#(y) = 1 * (3.0 - 1.0) + 2 * (3.0 - 2.0) + 3 * (2.5 - 0.5)$, where \( f_A(x,y) \) is the function evaluating the tensor \( A_{x,y} \). Another Einsum, \( s =  A_{x,y} * d(x,y;\mu_{\lambda},\mu_{\lambda}) \), yields zero because the 2D $xy$ area over a 1D $x$ interval is zero. An easy analogy is that the measure $\mu_{\lambda}$ acts as an integral over intervals, while the measure $\mu_{\#}$ resembles a discrete sum over points.

\subsubsection{\textbf{Boolean Semiring $(\{T,F\}, \vee , \wedge, F, T )$}} 

For logical disjunction, we use a function \( f_P: (x_1, \dots, x_n, rx_1, \dots, rx_k) \to \{\text{F}, \text{T}\} \). The reduction is defined by the logical measure $\mu_{\land\lor}$ :
\[
C_{x_1, \dots, x_n} = \bigvee_{rx_1 \dots rx_k} f_P(x_1, \dots, x_n, rx_1, \dots, rx_k) \, \land  \, \mu_{\land\lor}(rx_1) \land \dots \, \mu_{\land\lor}(rx_k)
\]
where \( C_{x_1, \dots, x_n} = T \) if \( f_P(x_1, \dots, x_n, rx_1, \dots, rx_k) = T \) for any $rx_1$ \dots $rx_k$ ; otherwise, \( C_{x_1, \dots, x_n} = F \). In this case, the boolean measure \( \mu_{\land\lor} \) assigns \( T \) to any non-empty set, and \( F \) to the empty set.

\color{black}
\section {Piecewise-Constant Specialization} \label{sec:piecewisespecialization}

Continuous tensor programs cannot reuse the operational definition from traditional Einsums~\cite{teaal}, as traversing all real-valued coordinates is not computable. To make our approach practical, we restrict our focus for the remainder of this paper to \textbf{piecewise-constant} tensors, where tensor values remain constant over specific intervals. This restriction enables efficient computation while still supporting a wide range of useful applications. In this subsection, we present a new evaluation semantics for continuous tensor programs under the piecewise-constant assumption. We then describe the validity conditions and explain how to verify whether a given program satisfies them. Finally, we discuss extensions of our system to support piecewise-non-constant functions.
\color{black}


\subsection{Piecewise-Constant Tensor} 
We introduce the piecewise-constant restriction to create a practical and implementable solution. In the rest of the paper, we assume: \textbf{\emph{All continuous tensors must be piecewise-constant.}}

We define a \textbf{tensor} \texttt{A} as a concrete data structure in memory whose value at real-valued indices \( x_1, \dots, x_n \) can be obtained by evaluating \texttt{A[$x_1, \dots, x_n$]}. We can express its values as a mathematical function \( f(x_1, \dots, x_n) = \texttt{A}[x_1, \dots, x_n] \). A tensor is said to be \textbf{piecewise constant} when its values are constant over specific regions of its domain and can be expressed as \( f(x_1, \dots, x_n) = \sum_{m} V_m \times \llbracket (x_1, \dots, x_n) \in I_m \rrbracket \), where: \( V_m \) is the constant value in the \( m \)-th piece; 
\( I_m \) is an \( N \)-dimensional interval (a hyperrectangle) representing the domain of the \( m \)-th piece; \( \llbracket  \rrbracket \) denotes the Iverson bracket, defined as \( \llbracket cond \rrbracket = 1 \) if the $cond$ is true, and \( 0 \) otherwise.

In this definition, the $N$-dimensional interval \( I_m \) must satisfy the following rules:

\begin{enumerate}[leftmargin=3em, label=\textbf{\arabic*.}]
    \item \textbf{Disjoint Intervals}: The intervals are pairwise disjoint, \( \forall\, m_1 \neq m_2, \ I_{m_1} \cap I_{m_2} = \emptyset \).
    \item \textbf{Complete Coverage}: The union of all intervals covers the entire domain, \( \bigcup_{m} I_m = \mathbb{R}^N \).
    \item \textbf{Axis-Aligned Intervals}: Each interval \( I_m \) is axis-aligned (i.e., no diagonal or curved shape).
\end{enumerate}

\begin{figure}
\centering
\includegraphics[width=\textwidth,valign=t]{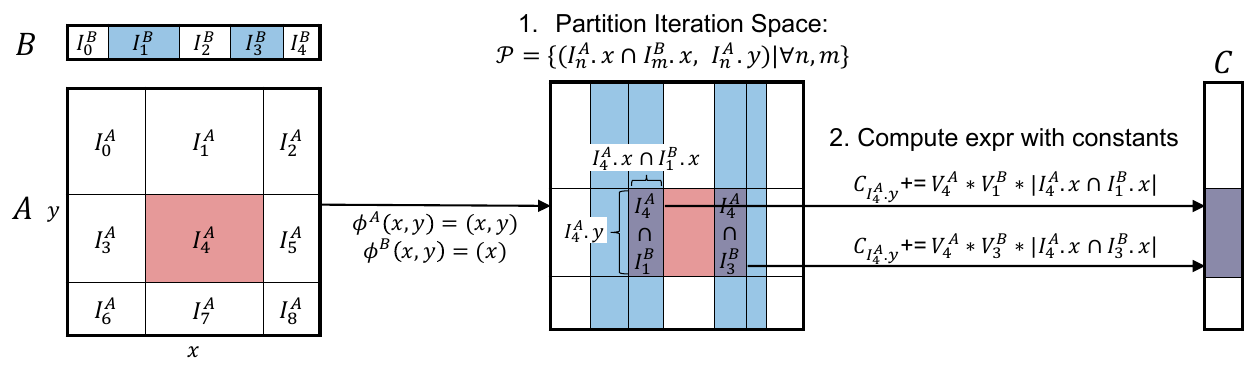}
\vspace{-0.2in}
\caption{Evaluation of \boldmath $ C_y = A_{x,y} * B_x * d(x,\mu_{\lambda}) $ \unboldmath, where $d(x,\mu_{\lambda})$ is a Lebesgue integral on \$x\$. The computation iterates over partitions formed by intersecting intervals of $A_{x,y}$ and $B_x$, with non-colored regions as zeros, and evaluates the expression within each. The two right arrows indicate computations on non-zero tensor elements, highlighting only interactions that contribute non-zero values.} 
\label{fig:opdef}
\end{figure}

\color{black}
\subsection{Piecewise-Constant Evaluation Semantics} 
\color{black}
In tensor computations involving piecewise constant tensors, we first partition the iteration space into regions where each tensor maintains a constant value. Rather than traversing all real-numbered coordinates in the entire space, we traverse each partition and evaluate the tensor expression using the constant values for that partition. We formalize an evaluation semantics as follows :

\textbf{1. Partitioning the Iteration Space.}
The goal is to partition the iteration space into regions where all tensors used in the expression remain constant. To achieve this, we first map piecewise-constant intervals of each tensor into the iteration space using indexing functions. An indexing function explains how each index is used to access the tensor. By applying the inverse of this indexing function, each tensor's constant interval is converted into a corresponding region in the iteration space (illustrated as Step 1a in Figure~\ref{fig:partitioning}). Next, the iteration space is partitioned by intersecting these regions from all tensors. Each resulting intersection indicates a unique region in which every tensor in the expression remains simultaneously constant. (illustrated as Step 1b)

\begin{figure}[b]
\centering
\begin{minipage}[t]{0.48\textwidth}
\centering
\begin{minted}[fontsize=\footnotesize,linenos,numbersep=1pt]{python}
# Step 1a: Map tensor's constant intervals into
# iteration space with inverse of indexing func
inverse_pieces = {}
for access in einsum: 
  tensor = access.tensor  
  indexingfunc = access.indexing 
  preimage = inverse(indexingfunc) 
  inverse_pieces[tensor] = [
      {'interval': preimage(piece.interval),
       'constant': piece.constant}
      for piece in tensor
  ]
\end{minted}
\vspace{-0.05in}
\caption*{(a) Step 1a}
\end{minipage}
\hspace{0.14in}
\begin{minipage}[t]{0.48\textwidth}
\centering
\begin{minted}[fontsize=\footnotesize,linenos,numbersep=1pt,firstnumber=13,escapeinside=||,mathescape=true]{python}
# Step 1b: Partition iteration space into regions
#          where tensors remain constant
for piece1 in inverse_pieces[tensor1]:
  ...
  for pieceN in inverse_pieces[tensorN]:
    intervals = [piece1.interval, ...,
                 pieceN.interval]
    partition = intersect(intervals)
    if partition.is_empty(): continue
    # Step 2: Evaluate expression using constants
    weight = 1
    if reduction exists: # if d(x,$\mu$) exists
      for (axis, measure) in einsum.d_exprs:
        weight |$\otimes$|= measure(partition.axis)
    out_constant = weight |$\otimes$| compute_expr(
      piece1.constant, ..., pieceN.constant)
    out_interval = out.indexing(partition)
    out[out_interval] |$\oplus$|= out_constant
\end{minted}
\vspace{-0.1in}
\caption*{\color{black}(b) Steps 1b \& 2\color{black}}
\end{minipage}
        \vspace{-0.1in}
    \caption{\color{black}Pseudocode of the piecewise-constant evaluation semantics of our language.\color{black}}
    \label{fig:partitioning}
\end{figure}

For instance, in Figure~\ref{fig:opdef}, the tensor expression \( C_y = A_{x,y} \ast B_x \ast d(x,\mu_{\lambda}) \) is partitioned as follows. The indexing functions are $\phi^{A}(x, y) = (x, y)$ and $\phi^{B}(x, y) = (x)$, with their respective pre-images:
\[
\phi^{A^{-1}}(I^{A}) = \{ (x, y) \in \mathcal{S} \mid (x, y) \in I^{A} \},  
\quad
\phi^{B^{-1}}(I^{B}) = \{ (x, y) \in \mathcal{S} \mid x \in I^{B} \},
\]
where $I^A$ and $I^B$ are intervals of tensor $A$ and $B$, and $\mathcal{S}$ denotes the iteration space of indices $(x,y)$. Note that the pre-image of an interval in tensor $B$ spans the entire region along the $y$-axis. The partitioned iteration space $\mathcal{P}$ is then defined as an intersection of the pre-image of:
\[
\mathcal{P} = \left\{ \phi^{A^{-1}}(I_{n}^{A}) \cap \phi^{B^{-1}}(I_{m}^{B}) \,\middle|\, \forall n, m \right\},
\]
where \( n \) and \( m \) range over all intervals of \( A \) and \( B \). In each partition, both \( A \) and \( B \) are constant.

\textbf{2. Compute Expression with Constants.} 
After partitioning, each disjoint partition contains constant values for all tensors involved in the expression. We traverse each partition and evaluate the expression using these constants. When the expression involves a reduction operation, we aggregate the computation results from each partition into the corresponding output region. If there is a reduction with Lebesgue measure over an axis, we multiply the result by the length of the partition along that axis to account for the integration over that interval. \color{black} Step 2 in Figure~\ref{fig:partitioning} illustrates this process. Note that the partition is an $n$-D axis-aligned rectangle \texttt{partition} $= I_1 \times I_2 \times \cdots \times I_n$, where $n$ is the number of indices in the expression. Let $R \subseteq \{1,\dots,n\}$ be the set of reduced axes, each with a semiring-valued measure $\mu_i : \mathrm{Intervals} \to \mathbb{S}$. In lines 25-26, the reduction contributes
$$
\texttt{out\_constant} \;\otimes\; \mu\left( \prod_{i \in R} I_i \right) = \texttt{out\_constant} \;\otimes\; \prod_{i \in R} \mu_i(I_i),
$$
which follows from the product measure for axis-aligned rectangles: by Fubini’s and Tonelli’s theorems~\cite{realanalysis}, the measure over the product equals the product of the individual measures.

To support an arbitrary semiring $(R, \oplus, \otimes, 0, 1)$ with arbitrary measures, users only need to implement lines 23-30 according to their chosen semiring and measure. To support a custom semiring, users initialize the output with the additive identity $0 \in R$, the measure weight with the multiplicative identity $1 \in R$, and define the implementations of $\oplus$ (addition) and $\otimes$ (multiplication). Users may also provide their own measure function $\mu : Interval \to R$ over intervals. For example, the \emph{Lebesgue measure} in the real-number semiring is defined as $\mu_\lambda([a,b]) = b - a$. As a concrete example, consider the \emph{min-product} semiring $(\mathbb{R}_{\ge 0} \cup \{\infty\}, \min, *, \infty, 1)$, where $\oplus = \min$ and $\otimes = *$. To use this semiring, one would initialize the output to $\infty$, and may implement an idempotent measure such as $\mu_{idem}([a,b]) = 1$, assigning uniform weight to all partitions regardless of size. 
\color{black}

In Figure~\ref{fig:opdef}, we illustrate this process by traversing all partitions. For illustrative purposes, we only depict effectual computations involving non-zeros. For each partition, we perform the following operation with constants: $C_{I^{\text{Partition}}_y} \mathrel{+}= V_n^A \times V_m^B \times \text{length}(I^{\text{Partition}}_x)$.


\color{black}
\subsection{Program Validity}
Not all programs are valid under the piecewise-constant assumption. To ensure correctness, the output must also be piecewise-constant—that is, the program must be \emph{closed under piecewise-constant assumption}. We establish the following conditions to guarantee this closure after partitioning:
\color{black}

\begin{enumerate}
    \item \textbf{Each \emph{effectual partition} must be an axis-aligned hyperrectangular interval.} 
    \item \textbf{Expressions within each effectual partition must remain constant, except when the partition reduces to a pinpoint.}    
\end{enumerate}

An \emph{effectual partition} refers to a partition where effectual computations—those that contribute to the final output, such as multiplying non-zero constants—take place. A \emph{pinpoint} refers to an N-D interval that collapses to a single point, meaning its end points are identical along all dimensions.

\begin{figure}
\centering
\begin{minipage}[t]{.25\textwidth}
\raisebox{-0.21in}{\includegraphics[width=\textwidth,valign=t]{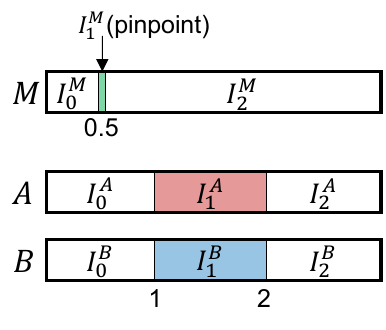}}
\vspace{-0.05in}
\subcaption{1D Continuous Tensors}
\label{fig:valida}
\end{minipage}
\hspace{0.04\textwidth}
\begin{minipage}[t]{.3\textwidth}
\centering
\raisebox{-0.08in}{\includegraphics[width=.82\textwidth,valign=t]{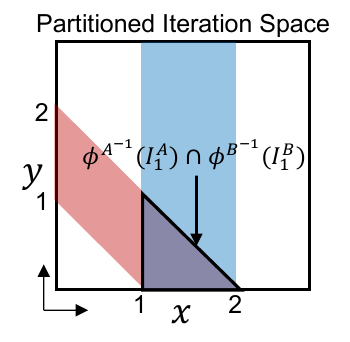}}
\vspace{-0.05in}
\subcaption{\centering Invalid Program:\\ \boldmath $C_y = A_{y+x} * B_x * d(x,\mu_{\lambda})$ \unboldmath}
\label{fig:validb}
\end{minipage}
\begin{minipage}[t]{.36\textwidth}
\centering
\raisebox{-0.06in}{\includegraphics[width=\textwidth,valign=t]{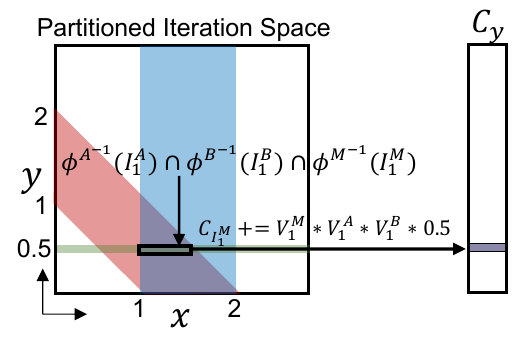}}
\vspace{-0.05in}
\subcaption{\centering Valid Program:\\ \boldmath $C_y = M_y * A_{y+x} * B_x * d(x,\mu_{\lambda})$ \unboldmath}
\label{fig:validc}
\end{minipage}
\vspace{-8pt}
\caption{Illustrations of partitioned iteration spaces and program validity. (a) shows 1D continuous tensors where non-colored regions have values of zero. (b) demonstrates an invalid program because the intersected partition is a triangular region, leading to a non-piecewise-constant output. (c) shows a valid program where the pinpoint $I_{1}^{M}$ in tensor $M$ adjusts the partition to form axis-aligned intervals.} 
\label{fig:valid}
\end{figure}

For the first criterion, the definition of a piecewise-constant tensor requires partitions to maintain an axis-aligned shape to ensure the output remains piecewise-constant. As illustrated in Figure~\ref{fig:validb}, a standard convolution on intervals creates triangular partitions, resulting in a non-constant output. As shown in Figure~\ref{fig:validc}, incorporating a pinpoint tensor \( M \) allows for the adjustment of effectual partitions into axis-aligned shapes, ensuring validity. This masking guarantees that the output tensor retains piecewise-constant property. Figure~\ref{fig:validity} shows an SMT-based validity-checking algorithm, commented with Figure~\ref{fig:validc}, that tests whether effectual partitions form hyperrectangles given tensor dimensions with nonzero pinpoints or intervals.

\begin{figure}[t]
\centering
\begin{minipage}{0.47\textwidth}
\centering
\begin{minted}[fontsize=\scriptsize]{python}
# Step 1: Construct a Symbolic Partition
# For example, in the expression 
# C[y] = M[y] * A[x + y] * B[x], 
# where M has pinpoints / A and B has intervals, 
# the partition over (x, y) satisfies :
# partition = [
#    y == p1,           # pinpoint M[y]       
#    p2 <= x + y <= p3, # interval A[x+y]
#    p4 <= x <= p5      # interval B[x]  
# ]
partition = []
indices = [x, y, z, ...]   # Index symbols
for access in einsum:        
  for tensordim in access:   
    indexExpr = tensordim.indexing(indices)  
    if tensordim.pinpoint:
      partition.add(
        indexExpr == FreshRealSymbol()
      ) 
    elif tensordim.interval:
      partition.add(
        FreshRealSymbol() <= indexExpr,
        indexExpr <= FreshRealSymbol()
      )
\end{minted}
\vspace{-0.05in}
\caption*{(a) Step 1 defines the shape of the partition by formulating linear inequalities and equalities of indices based on tensor accesses.}
\vspace{-0.05in}
\end{minipage}
\hfill
\begin{minipage}{0.48\textwidth}
\centering
\begin{minted}[fontsize=\scriptsize]{python}
# Step 2: Construct a Symbolic Hyperrectangle
# In the example:
# hyperrectangle = [
#    p6 <= x <= p7,               
#    p8 <= y <= p9,
# ]
hyperrectangle = []
for index in indices:
  hyperrectangle.add(
    FreshRealSymbol() <= index,
    index <= FreshRealSymbol()
  )

# Step 3: Check whether the partition 
#         always forms a hyperrectangle.
#
# In the example:
# Z3.check(ForAll([p1, p2, p3, p4, p5], 
#          Exists([p6, p7, p8, p9], 
#          ForAll([x, y], partition == hyperrectangle))))
Check1 = Exists(hyperrectangle.RealSymbols, 
         ForAll(indices, partition == hyperrectangle))
Check2 = ForAll(partition.RealSymbols, Check1)
Z3.check(Check2)
\end{minted}
\vspace{-0.05in}
\caption*{(b) Step 2 defines the hyperrectangular region over indices, and Step 3 checks whether the partition always forms a hyperrectangular shape.}
\end{minipage}
\vspace{-0.05in}
    \caption{An algorithm for verifying hyperrectangular partitions based on pinpoint/interval specifications.}
    \label{fig:validity}
\end{figure}

For the second criterion, expressions within each partition must remain constant to ensure that the computation remains constant throughout the partition. For example, the program \( C_x = A_x * (x+1) \) with a piecewise-constant vector \( A \) is invalid because the term \( (x+1) \) varies within any partition unless all effectual partitions are pinpoints. This check is only required for index expressions outside of tensor access. Such expressions are allowed only when the partition is guaranteed to be a pinpoint. To verify this, we use the same code structure as the hyperrectangularity check, but replace the hyperrectangle shape with a pinpoint to determine whether the partition is pinpoint.

\color{black}

\subsection{Arbitrary Piecewise Functions Beyond Constants}
Our piecewise‑constant abstraction sits at one extreme of a spectrum of richer piecewise function classes—polynomials, rationals, splines, and beyond. To extend our compiler to any such piecewise‑X class, we reuse the same partitioning logic from piecewise-constant evaluation semantics, but each tensor piece carries a symbolic description (e.g., polynomial coefficients) rather than a single scalar. In the constant case, the reduction over a product measure admits the shortcut $c \otimes\prod_i \mu_i(I_i)$ but with non‑constant integrands the compiler must instead evaluate the full nested integral: $\int_{I_1}\cdots\int_{I_n} g \;d\mu_n\cdots d\mu_1$ using closed‑form anti‑derivatives when available. Crucially, any chosen function class must be closed under every operations in piecewise evaluation semantics:
\begin{enumerate}
    \item \textbf{Algebraic operations}: Adding or multiplying two pieces must stay within the class.
    \item \textbf{Integration}: Integrating over any subinterval must either admit a closed‑form solution or be handled by a well‑defined approximation strategy.
\end{enumerate}

Polynomials satisfy all properties exactly. Most other classes fail at least one step—for example, rational integrals introduce logarithms, transcendental functions often lack closed forms, and fixed‑degree splines increase in degree under multiplication. In those cases, the compiler must detect the loss of closure, apply targeted approximation (e.g., numeric quadrature or projection back into the class), and propagate rigorous error bounds to guarantee overall accuracy. Supporting fully general piecewise‑X functions requires careful design with incorporating a symbolic engine like SymPy~\cite{sympy}, and is left to future work.

\subsection{Implementation}

\begin{figure}[t]
  \centering
  \includegraphics[width=\linewidth]{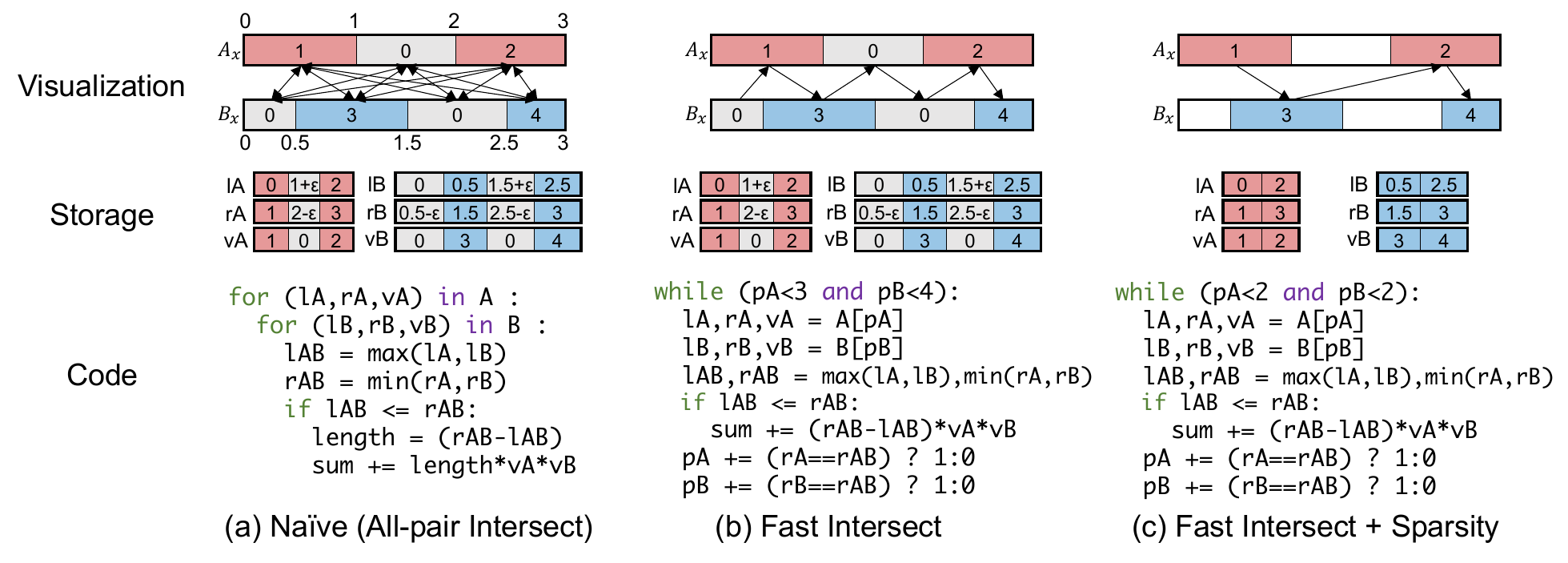}
  \vspace{-0.3in}
  \caption{Three strategies for partitioning the iteration space for \boldmath \( sum = A_x * B_x * d(x,\mu_{\lambda}) \) \unboldmath. \texttt{lT}, \texttt{rT}, and \texttt{vT} represent the left endpoints, right endpoints, and values of the interval of the tensor $T$, respectively. \textbf{(a) Naïve approach}: Computes all intersections using exhaustive pairwise comparisons. \textbf{(b) Fast Intersect}: Speeds up intersection using a two-pointer technique on sorted arrays. \textbf{(c) Fast Intersect + Sparsity}: Further improves (b) by leveraging sparsity to skip ineffectual partitions. Our compiler generates this final code.
  }
  \label{fig:gencode}
\end{figure}

While the piecewise-constant evaluation semantics describe partitioning the iteration space via all-pair intersections, we found that this process can be significantly optimized in practice. Figure~\ref{fig:gencode} shows two key optimizations which make our generated code more efficient. First, when partitioning the iteration space, it is unnecessary to compute all pairwise intersections between every interval of each tensor. Second, we can skip computations for partitions where ineffectual operations occur, such as \( a \times 0 = 0 \), by only storing nonzero intervals. Section~\ref{section:codegen} details how we generate efficient code to partition the iteration space and exploit sparsity using the Looplets \cite{finch} abstraction.

\color{black}

\section{Continuous Tensor Storage} \label{section:storage}
\begin{figure}
\centering
\includegraphics[width=\textwidth,valign=t]{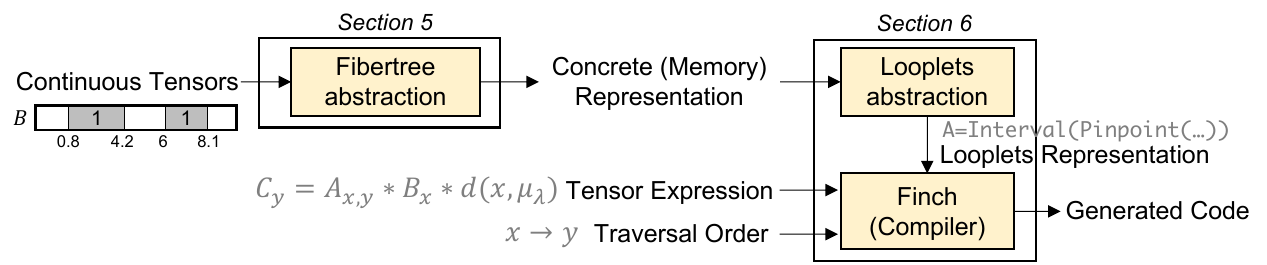}
\vspace{-0.15in}
\caption{Overview of our compiler implementation. The generated code efficiently partitions the iteration space into constant regions and computes the tensor expression within each partition. } 
\label{fig:imploverview}
\end{figure}

\color{black}
Our compiler implementation partitions the iteration space into constant regions efficiently and computes expressions within each constant partition, leveraging the fibertree~\cite{fibertree} and Looplets~\cite{finch}. Figure~\ref{fig:imploverview} provides an overview of our compiler implementation. This section will explain how the piecewise-constant continuous tensor is stored in memory using a fibertree abstraction. 
\color{black}

\subsection{Background on Format Abstraction}
This subsection provides background information on a format abstraction that provides an approach to storing traditional dense/sparse tensors in memory. Some of these abstractions are grounded in the coordinate tree concept, which was initially introduced in the context of the format abstraction~\cite{format} within TACO and subsequently refined and formalized as the \emph{fibertree} abstraction~\cite{fibertree}.

\begin{figure}
  \centering
  \includegraphics[width=\linewidth]{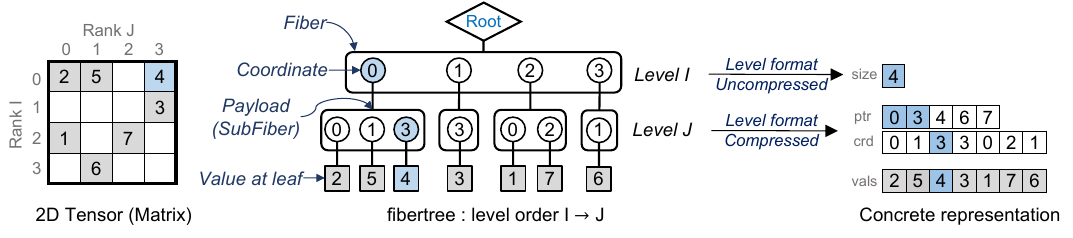}
  \vspace{-0.3in}
  \caption{A traditional sparse matrix, its fibertree abstraction and concrete representation stored in memory.}
  \label{fig:format}
\end{figure}

Figure~\ref{fig:format} illustrates how the fibertree abstraction depicts a 2D matrix. A \emph{tensor} is characterized as a multidimensional array with N \emph{ranks} (dimensions). The fibertree abstraction envisions a tensor as a tree structure, where each \emph{level} corresponds to a specific rank in the tensor. Each level comprises one or more fibers, representing sets of elements that share coordinates in the higher levels of the tree. Elements in a fiber are coordinate/\emph{payload} pairs, with the payload taking the form of either a (sub)fiber at the next level or a value located at the leaf of the tree. 

A \emph{level format} defines the physical storage used to store the fibers at that level. The two most prevalent level formats for integer coordinates are the \emph{Uncompressed} and \emph{Compressed} level formats. The Uncompressed level format encodes a dense integer coordinates within the range of [0, N). In contrast, the Compressed level format exclusively encodes non-zero integer coordinates within the fiber by explicitly storing their coordinates with offset pointers. In Figure~\ref{fig:format} (right), the concrete representation denotes that the matrix is stored in the $I \rightarrow J$ layout (row-major), with level formats assigned as Uncompressed for $I$ and Compressed for $J$. This specific representation corresponds to the Compressed Sparse Row (CSR) format. For more details on fibertrees, refer to~\cite{teaal, fibertree}.



\subsection{Interval Coordinates}


\begin{figure}
  \centering
  \includegraphics[width=.9\linewidth]{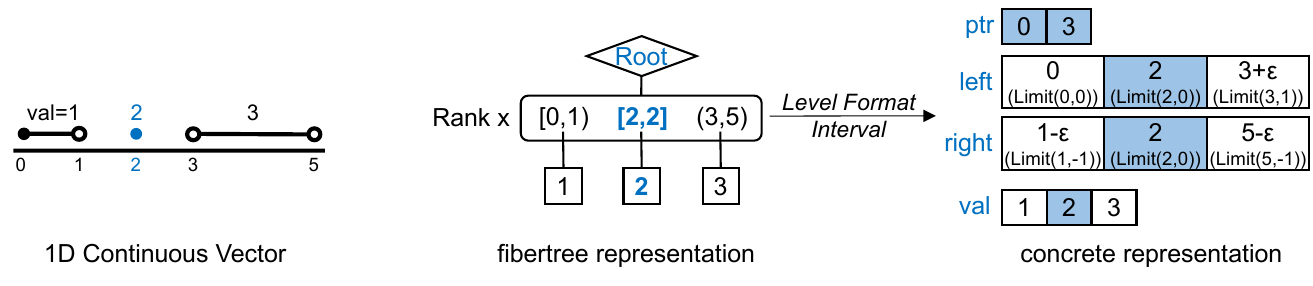}
  \vspace{-0.2in}
  \caption{A 1D continuous vector, its interval-typed fibertree abstraction and concrete representation.}
  \label{fig:cformat}
\end{figure}

We introduce interval-typed coordinates in a fibertree to effectively capture piecewise-constant properties.  Figure~\ref{fig:cformat} illustrates an interval-typed fibertree for a continuous tensor \texttt{A}. \emph{Pinpoint coordinates} are a special case, treated as intervals with both endpoints equal, collapsing to a single point; a pinpoint coordinate \texttt{2} is equivalent to the closed interval \texttt{[2,2]}. We account for the inclusiveness (open or closed) of each endpoint, resulting in four distinct interval subtypes. Similar to how traditional level formats operate on fibers with integer coordinates, we designed several level formats with interval coordinates, with the \emph{Interval level format} being assigned for rank $x$ in this example. Interval level format encodes the number and inclusiveness pair with \texttt{Limit} type.

\subsubsection{Interval Level Format with \texttt{Limit} Type.} 

To represent the interval inclusiveness (open and closed endpoints), we introduce a new number type called \texttt{Limit} for encoding interval endpoints. The \texttt{Limit} type incorporates an infinitesimal number, denoted by $\epsilon$, which is smaller than any positive real number but not zero. This approach allows us to treat all intervals as closed by default. We emulate open endpoints by adding or subtracting $\epsilon$ as needed, transforming open intervals into closed ones. Table~\ref{tbl:inclusiveness} illustrates how this representation works for each inclusiveness category.

\begin{table}[t]
\caption{Four inclusiveness types of intervals. In our implementation, we treat these as a single closed interval representation with the use of the infinitesimal number $\epsilon$ stored in \texttt{Limit} type.}
\vspace{-0.1in}
\footnotesize
\scalebox{0.9}{
\begin{tabular}{|l|l|l|l|l|}
\hline
\textbf{Name} & \textbf{Notation} & \textbf{Definition} & \textbf{Treated as} & \textbf{Implemented as} \\ \hline
Closed interval  & $ [a,b] $ & $ \{x\in \mathbb{R} \:|\: a \leq x \leq b\} $  & $[a, b]$ & \texttt{[Limit(a,0), Limit(b,0)]} \\
Right half-open interval & $ [a,b) $ & $  \{x\in \mathbb{R} \:|\: a \leq x < b\} $ & $[a, b-\epsilon]$ & \texttt{[Limit(a,0), Limit(b,-1)]} \\
Left half-open interval & $ (a,b] $ & $ \{x\in \mathbb{R} \:|\: a < x \leq b\} $ & $[a+\epsilon, b]$ & \texttt{[Limit(a,+1), Limit(b,0)]}\\
Open interval & $ (a,b) $ & $ \{x\in \mathbb{R} \:|\: a < x < b\} $ & $[a+\epsilon, b-\epsilon]$ & \texttt{[Limit(a,+1), Limit(b,-1)]}\\
\hline
\end{tabular}
}
\label{tbl:inclusiveness}
\end{table}

\begin{figure}
\begin{minipage}[t]{.8\textwidth}
\begin{minted}[fontsize=\scriptsize,frame=single, bgcolor=LightGray, framesep=1mm, linenos, numbersep=1pt]{octave}
# Definition of Limit.
struct Limit<T>
    val::T    # Numeric type T (e.g., Int, Float32) chosen by the user
    eps::Int8 # (+ε, 0, -ε) = (+1, 0, -1)
end
# Example Operations defined on Limit. 
(+)(x::Limit, y::Limit)::Limit = Limit(x.val + y.val, min(max(x.eps + y.eps, -1), +1))
(-)(x::Limit, y::Limit)::Limit = Limit(x.val - y.val, min(max(x.eps - y.eps, -1), +1))
(<)(x::Limit, y::Limit)::Bool = x.val < y.val || (x.val == y.val && x.eps < y.eps)
\end{minted}
\end{minipage}
\vspace{-0.2in}
\caption{Implementation of \texttt{Limit} type. Infinitesimal number is represented by \texttt{eps} field stored in \texttt{Int8} type.}
\label{fig:limit}
\end{figure}

Figure~\ref{fig:limit} depicts the definition of \texttt{Limit} and arithmetic operations implemented using this numeric type. \texttt{Limit} serves as an augmented number type for real numbers, essentially a struct comprising a regular numerical value (\texttt{val::T}) and the infinitesimal value (\texttt{eps::Int8}).  

Although our compiler generates symbolic code that operates on real coordinates or intervals, actual computations are executed using finite-precision computer number types. This finite precision inevitably introduces representation and rounding errors, distinguishing practical computations from idealized continuous real-number arithmetic.
The \texttt{Limit<T>} type supports various numeric types \texttt{T} (e.g., \texttt{Float32}, \texttt{Rational}, or \texttt{BigFloat} for arbitrary precision), allowing users to select the precision level that best fits their application’s requirements. For consistent comparison, we matched the numeric type for real indices to each baseline in our case studies (Section~\ref{sec:casestudy}).

\subsection{Optimized Representation}

This subsection outlines optimized representations for storing tensors with specific interval patterns. Choosing appropriate level order and level formats results in notable benefits in terms of memory efficiency and the complexity of the generated code.

\subsubsection{Impact of Level Order}

\begin{figure}[t]
  \centering
    \begin{minipage}[t]{.2\textwidth}
    \raisebox{-0.07in}{\includegraphics[width=\textwidth,valign=t]{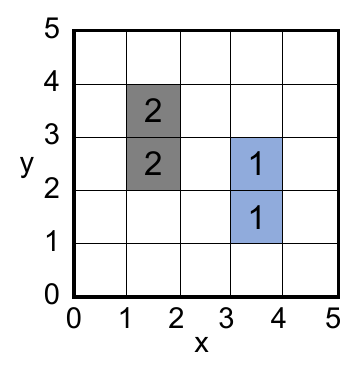}}
    \vspace{-0.05in}
    \subcaption{Continuous tensor}
    \label{fig:2dvis}
    \end{minipage}
    \hspace{0.0\textwidth}
    \begin{minipage}[t]{.34\textwidth}
    \includegraphics[width=\textwidth,valign=t]{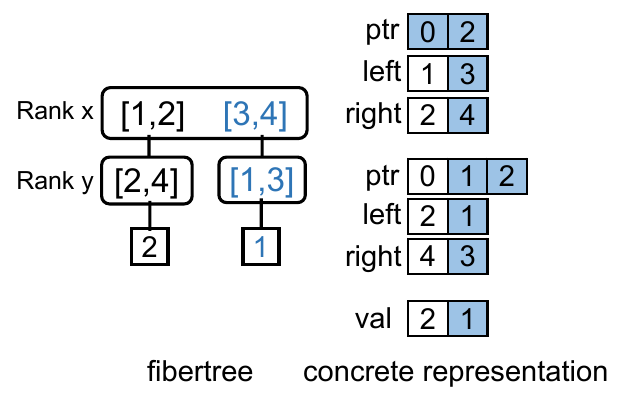}
    \vspace{-0.05in}
    \subcaption{Level order: $x \rightarrow y$ }
    \label{fig:2dfiber}
    \end{minipage}
    \hspace{0.0\textwidth}
    \begin{minipage}[t]{.42\textwidth}
    \includegraphics[width=\textwidth,valign=t]{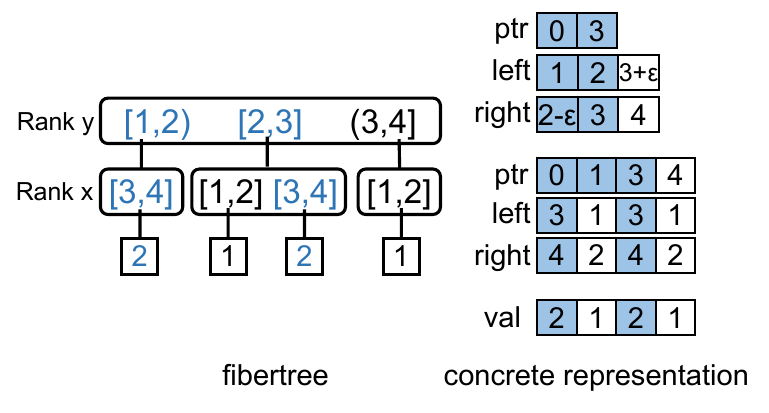}
    \vspace{-0.05in}
    \subcaption{Level order: $y \rightarrow x$ }
    \label{fig:overlapconcrete}
    \end{minipage}
    \hspace{0.0\textwidth}
    \vspace{-8pt}
  \caption{Two concrete representations based on different level orders. With the level order $x \rightarrow y$ and $y \rightarrow x$, the alternative representation illustrates how memory usage can be reduced depending on the chosen order.}
  \label{fig:lvlorder}
\end{figure}

The careful selection of the level order is crucial for optimizing tensor operations in both memory consumption and performance~\cite{waco}. As shown in Figure~\ref{fig:lvlorder}, the level order directly affects the structure of the fibertree and concrete representation. For instance, the order $x \rightarrow y$ and $y \rightarrow x$ results in different layouts. An optimal level order reduces redundancy in the storage format, minimizing memory usage by eliminating unnecessary index pointers. It also improves computation efficiency by reducing the overhead of irregular memory accesses.

\begin{figure}[t]
  \centering
  \includegraphics[width=0.8\textwidth]{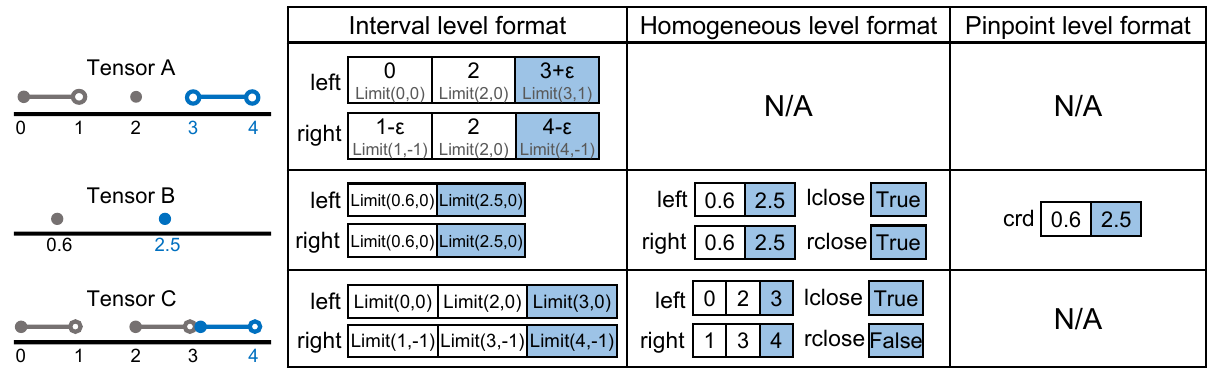}
    \vspace{-0.1in}
  \caption{Level formats across three distinct 1D continuous tensors. Instead of storing every pair of (number, inclusiveness) in the \texttt{Limit} type, certain tensors benefit from optimized level formats.}
  \label{fig:optrepr}
\end{figure}

\subsubsection{Impact of Level Format}
Figure~\ref{fig:optrepr} shows three tensors stored in various level formats. Depending on the pattern, certain tensors can benefit from a more optimized representation than storing every endpoint in the \texttt{Limit} type.

\para{Homogeneous Level Format}
The need to store pairs of (number, inclusiveness) for every endpoint may vary, depending on interval inclusiveness. When all intervals within a level share the same inclusiveness, we refer to them as \emph{homogeneous} intervals. Conversely, when intervals have inconsistent inclusiveness, they are categorized as \emph{heterogeneous}. Tensor B and C in Figure~\ref{fig:optrepr} are homogeneous but tensor A is heterogeneous. In the case of homogeneous intervals, there is no need to store every endpoint in the \texttt{Limit} type. Instead, the level format can store endpoints in regular numeric types while keeping inclusiveness details separate using \texttt{lclose} and \texttt{rclose}.

\para{Pinpoint Level Format}
If the level contains only pinpoint coordinates, there is no need to store both endpoints for each coordinate, as they share the same endpoints. Instead, pinpoints can be stored as single coordinates using the regular number type, as depicted in Tensor B in Figure~\ref{fig:optrepr}.

\section{Code Generation} \label{section:codegen}






In this section, we describe how our compiler transforms continuous loops into efficient, executable code by extending Finch~\cite{finchoopsla}, which was originally designed for sparse tensor computations in the integer domain using Looplets~\cite{finch}. We leverage Looplets to efficiently partition the iteration space, extending them into the continuous domain. At each step, repeated rewriting and optimization passes incorporate mathematical properties, such as sparsity, to enhance efficiency.

Our compiler takes two key inputs: (1) a continuous Einsum expression with index ordering, and (2) Looplet descriptions for each tensor. Using these inputs, the compiler generates executable Julia code. A key aspect of the compiler is its mechanism for lowering Looplets into loops.

\subsection{Background on Looplets}
\begin{figure}[t]
  \centering
  \includegraphics[width=\textwidth]{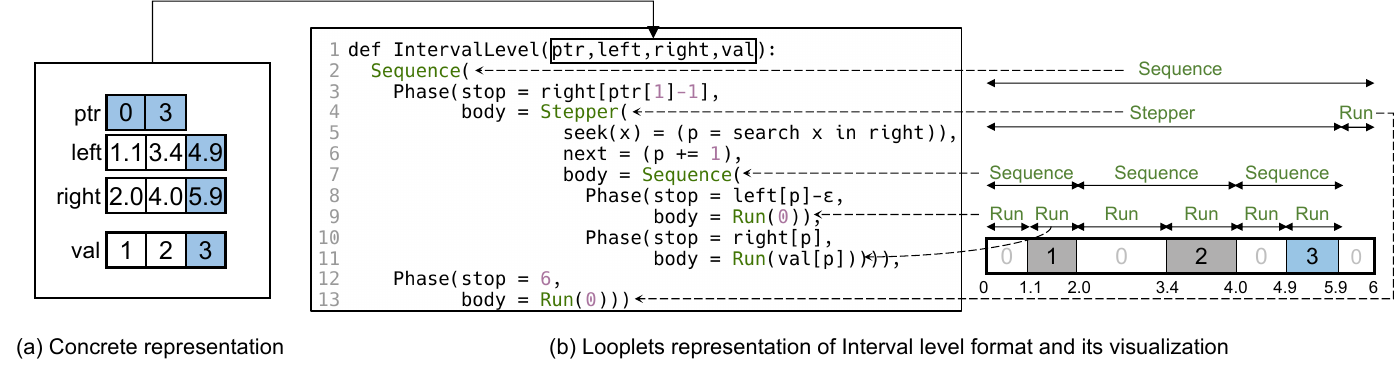}
  \vspace{-0.3in}
  \caption{(a) Concrete representation derived from fibertree using Interval level format. (b) Looplets representation describing the pattern of Interval level format using concrete representations. Here, $\epsilon$ is \texttt{Limit(0,1)}.  }
  \label{fig:Loopletrepr}
\end{figure}

While fibertrees describe tensors as a tree of levels, Looplets describe each fiber in a level as a tree of ranges, enabling more detailed and structured access patterns. The hierarchical decomposition provided by Looplets allows the compiler to resolve interactions, such as intersections or unions, between different structural formats, such as pinpoint levels and interval levels. Originally developed in the integer domain, but we found that Looplets can be extended naturally to the continuous domain with minimal modification. The following provides background information on four core types of Looplets designed for describing value patterns:

\begin{itemize} 
\item \textbf{Run}: Describes repeated instances of the same payload, scalar or sub-fiber.
\item \textbf{Sequence}: Concatenate multiple child Looplets into one sequence.
\item \textbf{Stepper}: Repeats a single child Looplet a variable number of times.
\item \textbf{Phase}: Marks the interval spanned by a child Looplet.
\end{itemize}

A useful analogy can be drawn between Looplets and regular expressions (Regex). A Run Looplet is similar to a single character in Regex, while a Sequence Looplet functions like concatenation (e.g., $RS$ for sub-Regex $R$ and $S$). A Stepper Looplet corresponds to the Kleene star ($R^*$), allowing for repeated patterns. For a comprehensive overview of Looplets, please refer to the original paper~\cite{finch}. 

Looplets provide a mechanism to interpret the concrete memory representations of tensors as full (which may include zeros) structures. Figure~\ref{fig:Loopletrepr} illustrates this process using an Interval Level format and how it can be mapped to a Looplet representation. In Figure~\ref{fig:Loopletrepr}(a), the concrete representation of  Interval level format is derived from fibertree. Figure~\ref{fig:Loopletrepr}(b) visualizes how this concrete Interval Level format is transformed into a Looplet representation. The Looplets representation decomposes the entire level into structured patterns. Inside the top Sequence, a Stepper Looplet iterates over the Sequence of zero and non-zero Runs. For non-zero Runs, the corresponding value (\texttt{val[p]}) is used, where \texttt{p} is the current position in the Stepper. Phases mark the boundaries of each interval using the \texttt{left} and \texttt{right} arrays. In this way, Looplets provide a hierarchical approach to understanding the concrete memory representation of continuous tensors.

\begin{figure}
\vspace{-0.15in}
\begin{minipage}[t]{.22\textwidth}
\raisebox{-0.1in}{\includegraphics[width=\textwidth,valign=t]{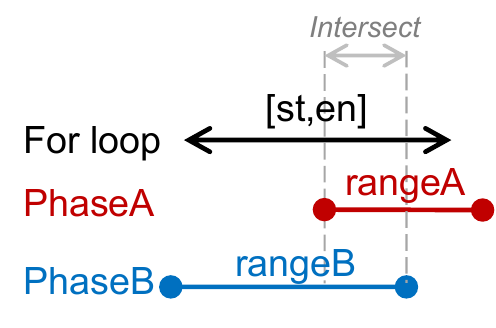}}
\subcaption{Phase: Visualization.}
\label{fig:codegena}
\end{minipage}
\hspace{0.05\textwidth}
\begin{minipage}[t]{.27\textwidth}
\begin{minted}[escapeinside=||,fontsize=\tiny,frame=single, bgcolor=LightGray, framesep=1mm]{octave}
# bodyA/B = child Looplet
A = Phase(rangeA, bodyA)
B = Phase(rangeB, bodyB)
|\textbf{\color{violet}{for x = st:en}}|
  |\textbf{\color{violet}{s += A[x] * B[x]}}|
\end{minted}
\subcaption{Phase: Source program.}
\label{fig:codegenb}
\end{minipage}
\hspace{0.05\textwidth}
\begin{minipage}[t]{.35\textwidth}
\begin{minted}[escapeinside=||,fontsize=\tiny,frame=single, bgcolor=LightGray, framesep=1mm, linenos, numbersep=1pt]{octave}
# A ∩ B = [max(A.start, B.start), 
#          min(A.stop, B.stop)]
intersect = [st,en] ∩ rangeA ∩ rangeB
if intersect.start <= intersect.stop
  |\textbf{\color{violet}{for x = intersect.start:intersect.stop}}|
    |\textbf{\color{violet}{s += bodyA[x] * bodyB[x]}}|
\end{minted}
\vspace{-0.25in}
\subcaption{Phase: Lowered program.}
\label{fig:codegenc}
\end{minipage}

\begin{minipage}[t]{.22\textwidth}
\raisebox{-0.1in}{\includegraphics[width=\textwidth,valign=t]{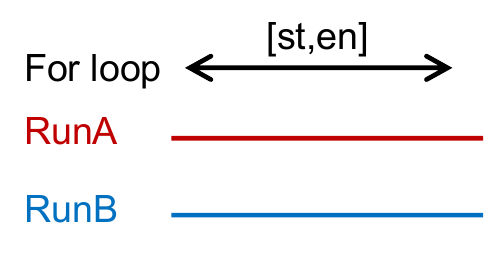}}
\vspace{-0.1in}
\subcaption{Run: Visualization}
\label{fig:codegend}
\end{minipage}
\hspace{0.05\textwidth}
\begin{minipage}[t]{.27\textwidth}
\begin{minted}[escapeinside=||,fontsize=\tiny,frame=single, bgcolor=LightGray, framesep=1mm]{octave}
A = Run(scalarA)
B = Run(scalarB)
|\textbf{\color{violet}{for x = st:en}}|
  |\textbf{\color{violet}{s += A[x] * B[x]}}|
\end{minted}
\vspace{-0.3in}
\subcaption{Run: Source program.}
\label{fig:codegene}
\end{minipage}
\hspace{0.05\textwidth}
\begin{minipage}[t]{.35\textwidth}
\begin{minted}[escapeinside=||,fontsize=\tiny,frame=single, bgcolor=LightGray, framesep=1mm, linenos, numbersep=1pt]{octave}
|\textbf{\color{violet}{for x = st:en}}|
  |\textbf{\color{violet}{s += scalarA * scalarB}}|
\end{minted}
\subcaption{Run: Lowered program.}
\label{fig:codegenf}
\end{minipage}

\begin{minipage}[t]{.22\textwidth}
\raisebox{-0.1in}{\includegraphics[width=\textwidth,valign=t]{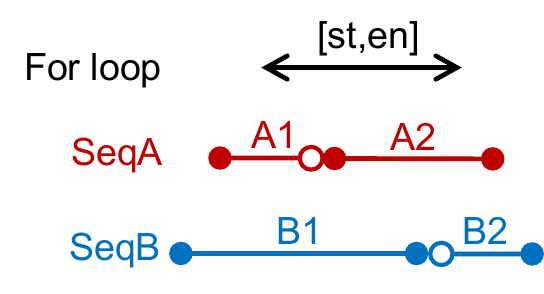}}
\subcaption{Seq: Visualization.}
\label{fig:codegeng}
\end{minipage}
\hspace{0.05\textwidth}
\begin{minipage}[t]{.27\textwidth}
\begin{minted}[escapeinside=||,fontsize=\tiny,frame=single, bgcolor=LightGray, framesep=1mm]{octave}
A = Sequence(PhaseA1, PhaseA2)
B = Sequence(PhaseB1, PhaseB2)
|\textbf{\color{violet}{for x = st:en}}|
  |\textbf{\color{violet}{s += A[x] * B[x]}}|
\end{minted}
\vspace{-0.2in}
\subcaption{Seq: Source program.}
\label{fig:codegenh}
\end{minipage}
\hspace{0.05\textwidth}
\begin{minipage}[t]{.35\textwidth}
\begin{minted}[escapeinside=||,fontsize=\tiny,frame=single, bgcolor=LightGray, framesep=1mm, linenos, numbersep=1pt]{octave}
|\textbf{\color{violet}{for x=st:en \{ s += PhaseA1[x]*PhaseB1[x] \}}}|
|\textbf{\color{violet}{for x=st:en \{ s += PhaseA1[x]*PhaseB2[x] \}}}|
|\textbf{\color{violet}{for x=st:en \{ s += PhaseA2[x]*PhaseB1[x] \}}}|
|\textbf{\color{violet}{for x=st:en \{ s += PhaseA2[x]*PhaseB2[x] \}}}|
\end{minted}
\vspace{-0.25in}
\subcaption{Sequence: Lowered program.}
\label{fig:codegeni}
\end{minipage}

\begin{minipage}[t]{.23\textwidth}
\raisebox{-0.1in}{\includegraphics[width=\textwidth,valign=t]{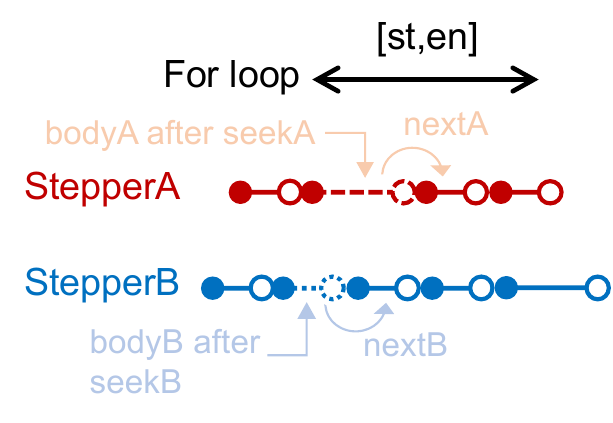}}
\subcaption{Stepper: Visualization.}
\label{fig:codegenj}
\end{minipage}
\hspace{0.05\textwidth}
\begin{minipage}[t]{.27\textwidth}
\begin{minted}[escapeinside=||,fontsize=\tiny,frame=single, bgcolor=LightGray, framesep=1mm]{octave}
A = Stepper(seekA,bodyA,nextA)
B = Stepper(seekB,bodyB,nextB)
|\textbf{\color{violet}{for x = st:en}}|
  |\textbf{\color{violet}{s += A[x] * B[x]}}|
\end{minted}
\subcaption{Stepper: Source program.}
\label{fig:codegenk}
\end{minipage}
\hspace{0.05\textwidth}
\begin{minipage}[t]{.37\textwidth}
\begin{minted}[escapeinside=||,fontsize=\tiny,frame=single, bgcolor=LightGray, framesep=1mm, linenos, numbersep=1pt]{octave}
call seekA(st) # find bodyA contains st
call seekB(st) # find bodyB contains st
x = st
while x <= en
  curr = [x,en] ∩ bodyA.range ∩ bodyB.range
  |\textbf{\color{violet}{for x2 = curr.start:curr.stop}}|
    |\textbf{\color{violet}{s += bodyA[x2] * bodyB[x2]}}|
  if (curr.stop==bodyA.stop) call nextA
  if (curr.stop==bodyB.stop) call nextB
  x = curr.stop + ε 
\end{minted}
\vspace{-0.25in}
\subcaption{Stepper: Lowered program.}
\label{fig:codegenl}
\end{minipage}
\vspace{-0.1in}
\caption{Compiler Pass for Looplets. Continuous loops are highlighted in purple. Non-purple code in right represents generated Julia code. The compiler recursively lowers nested Looplets until no purple code remains.}
\label{fig:codegen}
\end{figure}

\subsection{Compiler Pass for Looplets} 

In Finch, each Looplet type defined within a \texttt{for} loop statement is lowered by a corresponding compiler pass. We found that most of the Looplet passes can be applied to continuous space without modification, with the exception of the Stepper. Figure~\ref{fig:codegen} provides background information through a visualization of how each pass lowers the Looplets within a continuous loop.

\textbf{Phase} (Figure~\ref{fig:codegena},~\ref{fig:codegenb}, and ~\ref{fig:codegenc}). This pass lowers continuous loops associated with Phase Looplets . It intersects Phase ranges with the loop's range, generating code to check for intersections with a length $\geq 0$ (Lines 3-4 in Figure~\ref{fig:codegenc}). 
    
\textbf{Run} (Figure~\ref{fig:codegend},~\ref{fig:codegene}, and ~\ref{fig:codegenf}). This pass lowers Run Looplets by replacing with a constant.
    
\textbf{Sequence} (Figure~\ref{fig:codegeng},~\ref{fig:codegenh}, and ~\ref{fig:codegeni}). This pass lowers the continuous loop associated with Sequence Looplets, concatenating multiple child Phases. This pass generates all combinations of child Phases within each Sequence: \texttt{(A1,B1), (A1,B2), (A2,B1), (A2,B2)}.

\textbf{Stepper} (Figure~\ref{fig:codegenj},~\ref{fig:codegenk}, and ~\ref{fig:codegenl}). This pass lowers Stepper Looplets, which repeat the child Looplet step by step. Each Stepper maintains a current body Looplet and performs computations on the intersected range. It then advances to the next body Looplet when the current one is complete.  The \texttt{seek} field contains code to fast-forward the stepper to the first body containing the start (\texttt{st}) of the for loop (Lines 1-2 in Figure~\ref{fig:codegenl}). Within the while loop, computations occur in the intersected range between current bodies (Lines 5-7). When the current body is complete, the Stepper advances to the next body (Lines 8-9). Finally, the next starting point \texttt{x} is set to the end of the intersection incremented by $\epsilon$ (\texttt{x = curr.stop + 1}). This $\epsilon$ is essentially implemented as \texttt{Limit(0, Int8(1))}. This increment changes a closed ending point to an open starting point, and vice-versa.

The process of compiling continuous loops into executable code entails a recursive lowering of these loops until none of them are present in the code (i.e., until there is no purple code). A specific compiler pass is selected based on the type of Looplet, with a focus on lowering the outermost first in nested Looplets. In scenarios where different tensors feature varying outermost Looplet types, tiebreaking rules establish the priority order as \texttt{Run > Phase > Sequence > Stepper}.

\pagebreak
\subsection{Compiler Pass for Simplifying Program}
\begin{wrapfigure}{l}{0.36\textwidth}
    \centering
    \vspace{-0.19in}
\begin{minted}[fontsize=\scriptsize, frame=single]{octave}
+(a..., 0, b...) => +(a..., b...)
*(a..., 0, b...) => 0
&&(a..., true, b...) => &&(a..., b...)
&&(a..., false, b...) => false
a[i...] += 0 => emptyblock()
a[i...] *= 1 => emptyblock()
a[i...] &= true => emptyblock()
a[i...] |= false => emptyblock()
if(true, a) => a
if(false, a) => emptyblock()
for x=a:b; emptyblock() => emptyblock()
\end{minted}
\vspace{-0.1in}
    \caption{Example Rewrite Rules in Finch}
    \label{fig:rewriting}
\vspace{-0.15in}
\end{wrapfigure}
\para{Background} Finch's simplify pass plays a crucial role in  optimization. It operates after each Looplet Pass, simplifying the program through predefined rewriting rules that account for mathematical properties. Examples of such rules are illustrated in Figure~\ref{fig:rewriting}. These rules extend beyond basic optimizations like constant propagation; some also impact control flow, such as \texttt{for} or \texttt{if} statements.While the existing rules were initially designed for the integer domain, they are equally applicable to continuous loops. \color{black} By leveraging annihilators in arbitrary semirings, Finch’s rewriting system can optimize continuous-domain programs as well. The current implementation supports diverse semirings such as the number, Boolean, tropical, min-max, min-product, and max-product semirings. Users can also extend the system to support custom semirings by adding new rules. \color{black}

\para{Additional Simplifications for Continuous Reduction} The main challenge in lowering continuous loops lies in managing infinite operations. However, with piecewise-constant tensors, we can simplify by focusing on constant regions, distinguishing reduction modes based on measure definition as described in Section~\ref{sec:piecewisespecialization}. Idempotent reductions like \texttt{max} and \texttt{min} are straightforward, as they behave consistently across both points and intervals. For the \texttt{+} reduction, we define two measures: summation (interpreted via the counting measure $\mu_\#$) and integration (interpreted via the Lebesgue measure $\mu_\lambda$). Summation aggregates over discrete points, while integration accumulates values across intervals by multiplying by interval length. 



\begin{figure}
\vspace{-0.1in}
\begin{minipage}[t]{.48\textwidth}
\begin{minted}[escapeinside=||, fontsize=\scriptsize,frame=single, bgcolor=LightGray]{octave}
# Rule for rewriting continuous reduction
rewrite_rule(for idx in interval; body) => begin
  apply_rule((op=)(lhs,rhs*d(idx,measure)) => begin
    if rhs is constant with respect to idx then
      reduce(idx, interval, lhs, op, rhs, measure)
    end
  end)(body)
end
\end{minted}
\vspace{-0.4in}
\begin{minted}[escapeinside=||, fontsize=\scriptsize,frame=single, bgcolor=LightGray]{octave}
# Limit(value, eps)
st = Limit(3.0,+1) #3.0 + ε 
en = Limit(4.2,0)  #4.2
|\textbf{\color{violet}{for x = st:en}}|
  |\textbf{\color{violet}{s += Va * Vb * d(x, Lebesgue)}}|
↓↓↓↓↓↓↓↓↓↓↓↓↓
s += Va * Vb * drop_eps(en-st) # s += Va*Vb*1.2

\end{minted}
\vspace{-0.2in}
\subcaption{Continuous reduction rule and an example.}
\label{fig:reductionb}
\end{minipage}
\hspace{0.02\textwidth}
\begin{minipage}[t]{.48\textwidth}
\begin{minted}[fontsize=\scriptsize,frame=single, bgcolor=LightGray]{octave}
# Function to drop epsilon (e.g., 3 + ε => 3)
drop_eps(x::Limit) = x.value

# Lebesgue measure
reduce(idx, interval, lhs, +=, rhs, Lebesgue) =
  interval_length = length(interval)
  lhs += rhs * drop_eps(interval_length)

# Counting measure
reduce(idx, interval, lhs, +=, rhs, Counting) =
  if length(interval) == 0 then
    lhs += rhs
  end

# Boolean measure
reduce(idx, interval, lhs, ||=, rhs, Boolean) =
  lhs = lhs || rhs 
\end{minted}
\vspace{-0.2in}
\subcaption{Collapsing terms based on operator and context.}
\label{fig:reductiona}
\end{minipage}
\vspace{-0.1in}
\caption{\color{black} Rewriting rule for continuous reduction. When the rule identifies that the assignment is reducible with respect to a loop, it substitutes the loop into the collapsed expression. \color{black} }
\label{fig:reduction}
\end{figure}

Continuous reduction is achieved through the addition of rewriting rules in the simplify pass, as depicted in Figure~\ref{fig:reductionb}. When this rule detects a \texttt{for} loop, it substitutes all applicable assignments into the collapsed expression, provided that the assignment is reducible with respect to the loop. \color{black} Figure~\ref{fig:reductiona} shows three collapsed expressions—two measures in the real-number semiring, and a boolean measure in the boolean semiring. As discussed in the piecewise-constant evaluation semantics, users can define custom measures $\mu\colon \text{Interval} \to R$ through rewrite rules.\color{black}

In Lebesgue measure, we utilize the \texttt{drop\_eps} function on the \texttt{Limit} type to remove epsilon from the length of the loop interval. In Figure~\ref{fig:reductiona}, \texttt{drop\_eps} extracts the number(\texttt{x.value}) from \texttt{Limit} type. This is done because integration with Lebesgue measure yields the same result regardless of the inclusiveness of the interval (i.e., $\int_{[0,1]} f(x)dx = \int_{(0,1)} f(x)dx$). Figure~\ref{fig:reductionb} below provides an example of how a continuous for loop is reduced using Lebesgue measure. In summation mode using counting measure, we emit an additional condition to check if the interval is pinpoint (i.e., the length of the interval is zero). This ensures that summation only operates on pinpoint pieces.

\para{Simplifying Interval Operations with Z3} Our compiler frequently generates code that handles interval operations, such as intersection checks or length computations. To optimize these operations, we utilize compile-time information about interval relationships within nested Looplets. We then employ the Z3 solver~\cite{z3} to prove certain interval conditions statically, allowing us to eliminate unnecessary computations. A typical example occurs when checking if the intersection between two intervals (\texttt{A} $\cap$ \texttt{B}) has zero length, especially when one interval is a pinpoint where \texttt{B.start == B.end}. Using Z3, we verify that pinpoint intersections always have zero length. Therefore, explicit length checks become unnecessary, reducing overhead and improving performance, especially in computations involving numerous pinpoint intersections.

\begin{figure}[H]
    \centering
    \begin{subfigure}[b]{0.48\textwidth}
        \centering
        \begin{minipage}{0.6\textwidth}
            \centering
            \begin{minted}[fontsize=\footnotesize]{python}
ABStart = min(A.end, B.end)
ABEnd = max(A.start, B.start)
if ABEnd >= ABStart:
    Length = ABEnd - ABStart
    if Length == 0:
        Sum += val
            \end{minted}
        \end{minipage}
        \caption{Original Code}
        \label{fig:original-pinpoint-intersection}
    \end{subfigure}
    \hfill
    \begin{subfigure}[b]{0.48\textwidth}
        \centering
        \begin{minipage}{.6\textwidth}
            \centering
            \begin{minted}[fontsize=\footnotesize]{python}
ABStart = min(A.end, B.end)
ABEnd = max(A.start, B.start)
if ABEnd >= ABStart:
    Sum += val
            \end{minted}
        \end{minipage}
        \caption{Simplified Code}
        \label{fig:simplified-pinpoint-intersection}
    \end{subfigure}
    \vspace{-0.1in}
    \caption{Optimizing pinpoint interval intersection checks using Z3.}
    \label{fig:pinpoint-optimization}
\end{figure}

\section{Case Studies} \label{sec:casestudy}
In this section, we explore diverse applications under the continuous tensor abstraction across four domains: (1) Geospatial search, (2) Genomic interval operations in Bioinformatics, (3) Interpolation in Neural Radiance Field, and (4) 3D point cloud convolution.  Our goal is to show the simplicity and clarity with which these applications can be expressed in our abstraction, particularly compared to challenges in existing tensor programming (Table~\ref{tbl:loc}). \color{black} Additionally, we report compilation time. Our extension for continuous tensors adds little overhead. Most of the compile time comes from the number of indices in Einsum, as each index is lowered into its own looplet, leading to deeper loop nests and more IR to generate and optimize. In practice this is a one‐time cost, since the compiled kernel can be reused for any inputs. \color{black} We assess the performance of the code generated by our compiler by comparing it with hand-written libraries. All experiments are conducted through single-threaded execution on a Macbook Pro M2 Max with 32GB of memory. 

\begin{table}[t]
\caption{Lines of code comparison between the baseline and ours, along with our compilation times.}
\vspace{-0.15in}
\footnotesize
\scalebox{0.95}{
\begin{tabular}{|l|l|l|l|l|}
\hline
\textbf{Applications}              & \textbf{Baseline} & \textbf{Ours} & \textbf{LoC Saving} & \textbf{Ours Compilation Time} \\ \hline
Radius Search Query                & 501 lines     & 8 lines    & \textbf{62$\times$}     & 16s \\ \hline
Genomic Interval Overlapping Query & 206 lines     & 11 lines   & \textbf{18$\times$}     & 18s \\ \hline
Trilinear Interpolation in NeRF    & 82 lines      & 13 lines   & \textbf{6$\times$}      & 28s \\ \hline
Point Cloud Convolution            & 2,330 lines   & 23 lines   & \textbf{101$\times$}    & 280s \\ \hline
\end{tabular}
}
\label{tbl:loc}
\end{table}

\subsection{Geospatial Search}

The first application we have explored is the spatial search query on 2D points, a widely used technique in applications such as geographical information systems (GIS)~\cite{gis}, computer-aided design (CAD)~\cite{cad}, and spatial databases~\cite{spatialdatabase}. In our study, we focused on two commonly employed queries: (1) the box search and (2) the radius search. In a box search, given a set of 2D points, this query retrieves all points within a specified box. A radius search query retrieves all points within a circle centered at ($O_x$, $O_y$) with a radius of $R$.

\begin{figure}
\centering
\begin{minipage}[t]{.50\textwidth}
\raisebox{0.1in}{\includegraphics[width=\textwidth,valign=t]{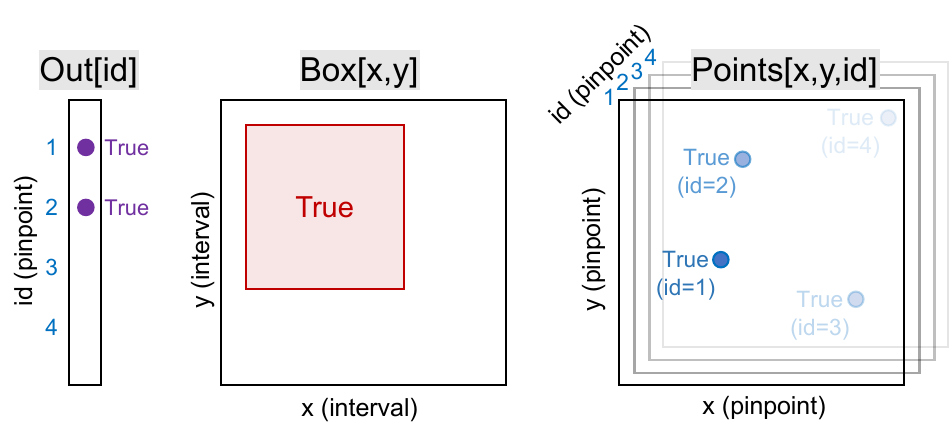}}
\vspace{-0.12in}
\subcaption{\centering \textbf{Box search query}:\\ \scalebox{0.9}{$Out_{id}$ = ($Box_{x,y}$ \&\& $Points_{x,y,id}$) * $d(x,y;\mu_{\land\lor},\mu_{\land\lor})$ }}
\label{fig:spatiala}
\end{minipage}
\centering
\begin{minipage}[t]{.45\textwidth}
{\includegraphics[width=\textwidth,valign=t]{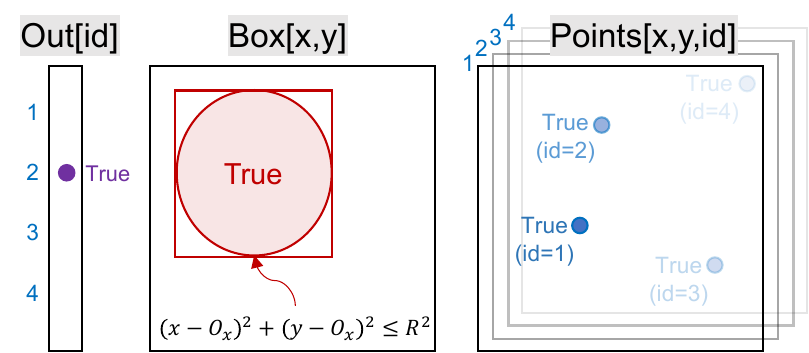}}
\subcaption{\centering\textbf{Radius search query}:\\ \scalebox{0.76}{$Out_{id}$ = $Box_{x,y}$ \&\& $Points_{x,y,id}$ \&\& $(x-O_x)^2+(y-O_y)^2\leq R^2$}}
\label{fig:spatialb}
\end{minipage}
\vspace{-0.1in}
\caption{Illustration of two spatial searches. Points are represented as a 3D tensor \texttt{Points[x, y, id]}, with each point assigned a unique ID. \texttt{Out} retrieves the \texttt{id} of points intersecting a specified box or circle. Measure on the radius search query omitted for brevity; it also uses boolean measures \(d(x, y; \mu_{\land\lor}, \mu_{\land\lor})\).}
\label{fig:spatial}
\end{figure}

\begin{figure}[t]
  \centering
  \captionsetup{justification=centering}
    \begin{minipage}[t]{.49\textwidth}
      \centering
    \includegraphics[width=0.8\textwidth,valign=t]{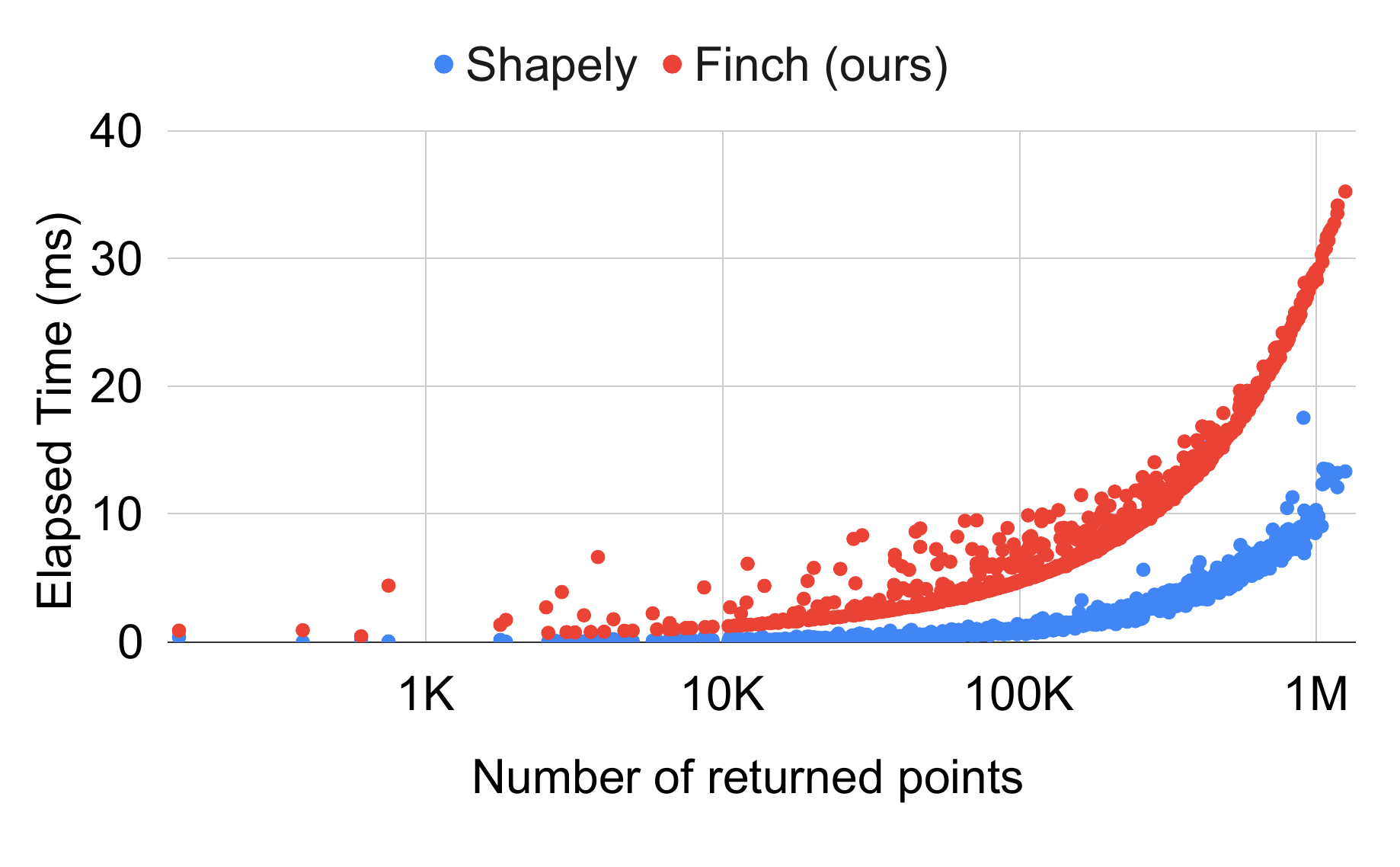}
    \vspace{-0.15in}
    \subcaption{Box search query with increasing box size.}
    \label{fig:boxeval}
    \end{minipage}
    \begin{minipage}[t]{.49\textwidth}
      \centering
    \includegraphics[width=0.8\textwidth,valign=t]{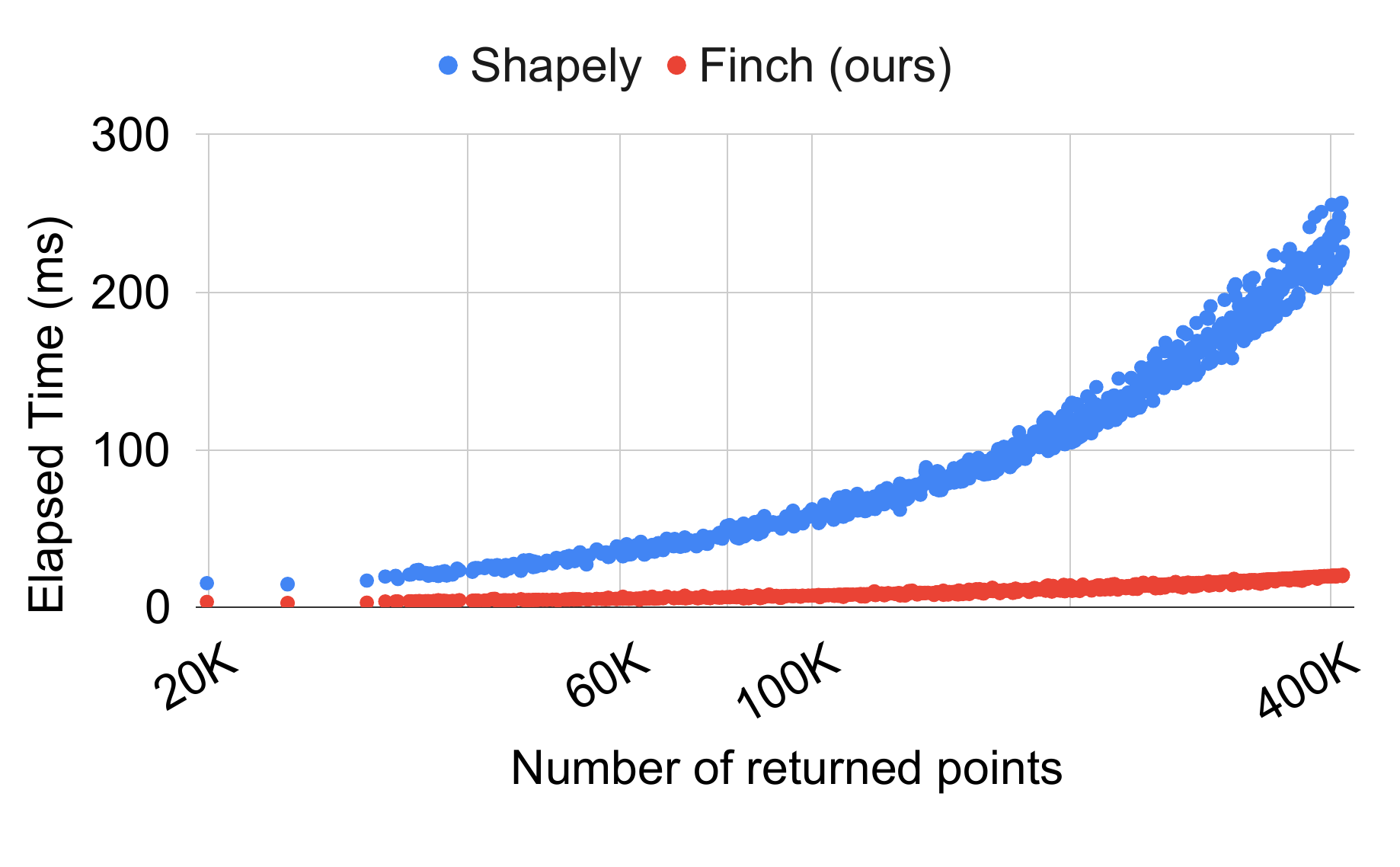}
    \vspace{-0.15in}
    \subcaption{Radius search query with increasing radius.}
    \label{fig:radiuseval}
    \end{minipage}
\vspace{-0.1in}
  \caption{Experimental result of spatial queries: lower values indicate better performance.}
  \label{fig:spatialeval}
\end{figure}

Figure~\ref{fig:spatial} demonstrates how spatial search queries are represented in continuous tensor abstraction. We transform 2D points into a 3D continuous tensor, denoted as $Points_{x,y,id}$, by assigning a unique ID to each point $(x, y)$. This approach accommodates multiple points that may share the same coordinates $(x, y)$ depending on the dataset. Figure~\ref{fig:spatiala} illustrates a box query, which outputs IDs of points intersecting with $Box_{x,y}$. Similarly, Figure~\ref{fig:spatialb} presents a radius search query where the circle is defined with $(x-O_x)^2+(y-O_y)^2 \leq R^2$.

Figure~\ref{fig:spatialeval} presents the performance of our generated code in comparison to Shapely~\cite{shapely}, a Python wrapper for GEOS~\cite{geos}, a well-known C++ library widely used in GIS for performing operations on two-dimensional geometries. We employed a synthetic dataset that uniformly distributed 10 million points in the range [0,10000] $\times$ [0,10000]. In both experiments, we increased the size of the query shape along the $x$-axis to augment the number of returned output points.

Figure~\ref{fig:boxeval} shows that Shapely outperforms our generated code on the box search query, achieving a geometric mean speedup of 4.7$\times$. This advantage is mainly due to Shapely's use of an advanced spatial data structure, STRtree~\cite{strtree}, which accelerates search operations. In contrast, our code does not employ any spatial data structures. However, Figure~\ref{fig:radiuseval} shows that our code surpasses Shapely on the radius query by a geometric mean of 9.2$\times$. This improvement ironically stems from Shapely's reliance on STRtree, which only supports box queries. For a circular query, Shapely retrieves all points within the bounding box of a circle using STRtree and then performs a linear scan to check if each point lies within the radius $R$. In contrast, our generated code directly iterates within the circular region (Figure~\ref{fig:spatialb}), avoiding the inefficiency of Shapely's two-step approach. 

\subsection{Genomic Interval Operations} \label{subsec:genomic}

\begin{figure}[t]
\raisebox{-0.2in}{\includegraphics[width=\textwidth,valign=t]{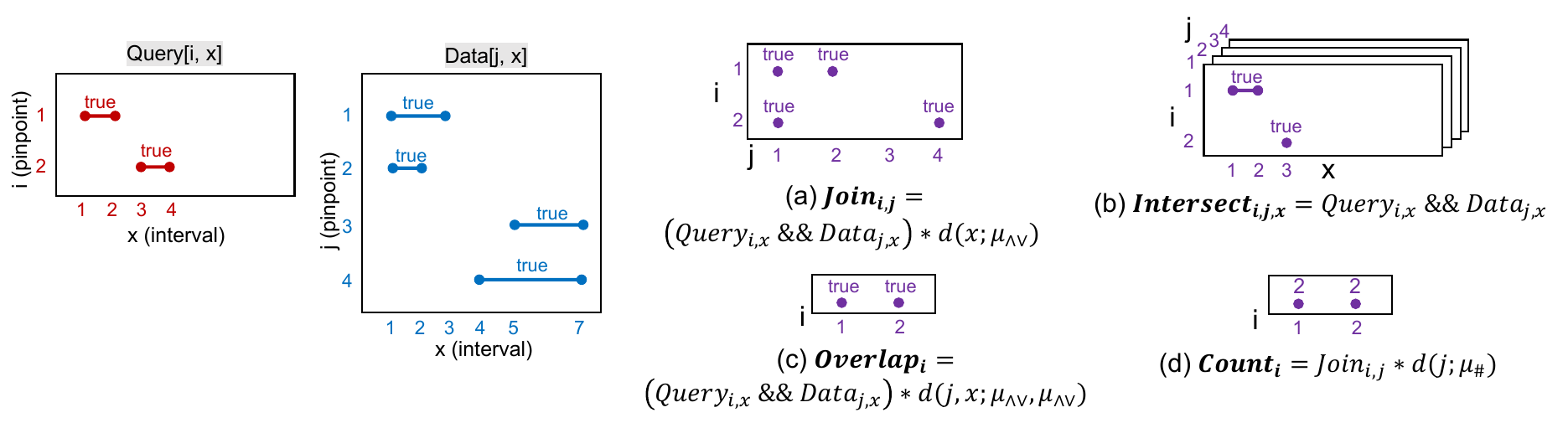}}
\vspace{-0.2in}
\caption{Four different genomic interval operations in continuous tensor abstraction. Genomic intervals are represented as 2D continuous tensor ($Query$ and $Data$) where white regions indicate 'false' boolean values. }
\label{fig:genomic}
\end{figure}

\begin{figure}[t]
\raisebox{-0.2in}{\includegraphics[width=\textwidth,valign=t]{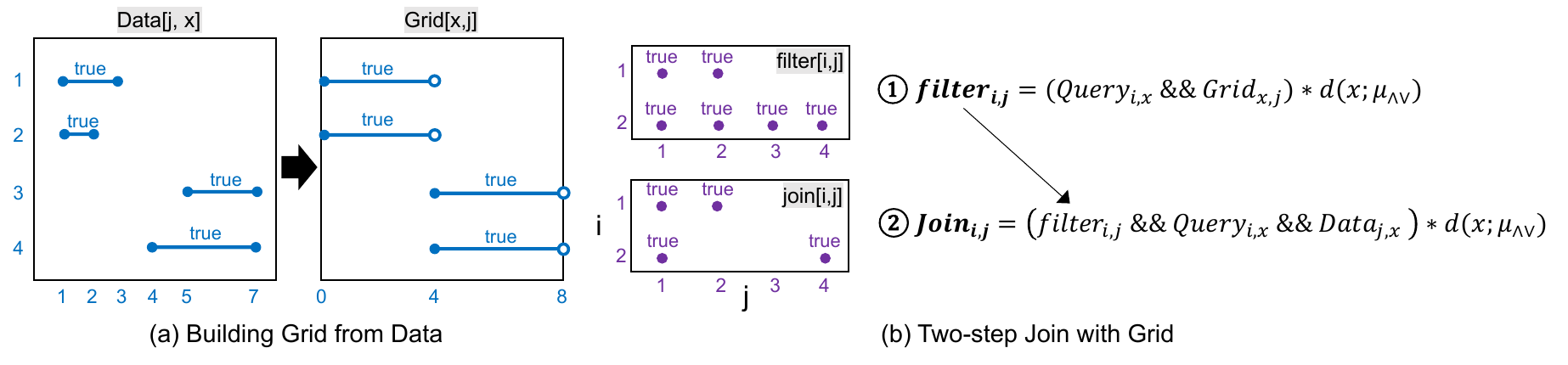}}
\vspace{-0.2in}
\caption{The uniform grid splits the $x$-dimension of `Data` into two intervals, \([0,4)\) and \([4,8)\), to streamline intersection tests by eliminating irrelevant data points. (a) illustrates the construction of the grid. (b) demonstrates a two-step join operation using the grid: the first step ($filter_{i,j}$) prunes unnecessary pairs \((i,j)\) with the grid structure, while the second step ($Join_{i,j}$) performs the actual join operation only on filtered pairs.}
\label{fig:grid}
\end{figure}

The second application we've explored is genomic interval operations using our continuous abstraction. In Bioinformatics, performing operations on genomic sequences~\cite{Bedtools} and applying boolean operations to genomic interval data can be computationally intensive. 

In Figure~\ref{fig:genomic}, genomic interval data is depicted in a 2D continuous tensor, denoted as the interval database $Data_{j,x}$ and query intervals $Query_{i,x}$, each identified by a unique interval ID. Right side of the figure illustrates four genomic operations written in the continuous tensor abstraction. For example, $Count_i$ counts the number of intervals in $Data$ intersecting with a query interval for each query ID $i$. These programs are expressed clearly and succinctly in the continuous tensor abstraction. However, the naive version involves pairwise comparisons ($N \times M$) between all query and data intervals ($i,j$), incurring substantial computational costs when handling numerous intervals.

In practice, previous studies~\cite{igd,ncls,ail,Bedtools,biofast} have used an interval data structure to skip unnecessary comparisons. We demonstrated a uniform grid~\cite{igd} using our abstraction to partition the continuous dimension $x$ into exclusive partitions, encompassing the entire dimension $x$. The uniform grid, represented in a 2D continuous tensor as $Grid_{x,j}$, is illustrated in Figure~\ref{fig:grid}. It divides the  $x$ domain of $Data_{j,x}$ into two halves: [0,4) and [4,8), with interval IDs $j$ allocated to the respective partitions. The right side of the figure shows an example on a Join operation using the uniform grid. First, the uniform grid filters out irrelevant intervals in Data, reducing the need for pairwise comparisons. In the second stage, it further eliminates false positives to refine the final results.

\begin{figure}[t]
  \centering
  \captionsetup{justification=centering}
    \begin{minipage}[t]{.49\textwidth}
    \includegraphics[width=\textwidth,valign=t]{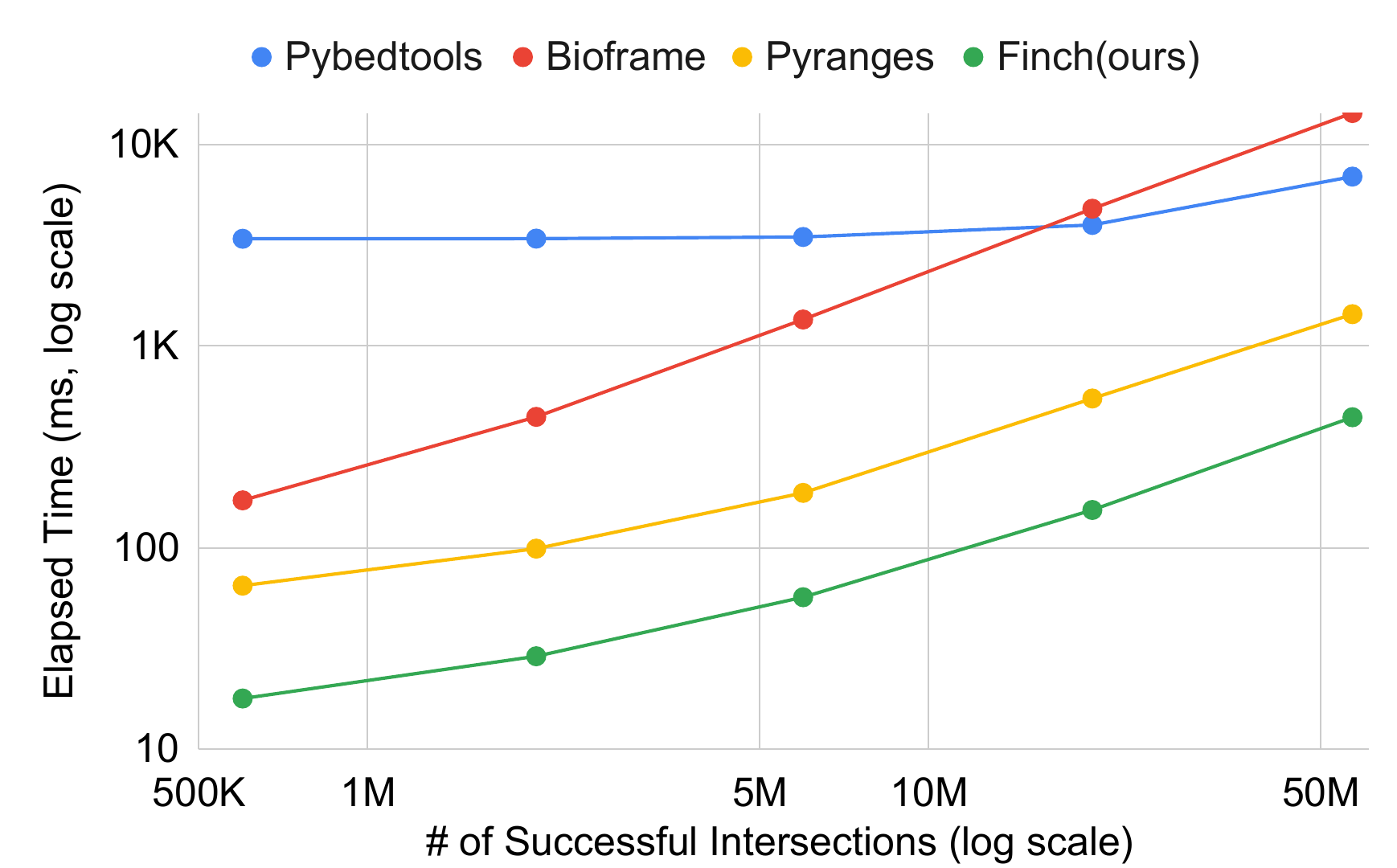}
    \subcaption{Sensitivity test regarding the count of successful intersections in a synthetic dataset.}
    \label{fig:genomiceval1}
    \end{minipage}
    \begin{minipage}[t]{.49\textwidth}
    \includegraphics[width=\textwidth,valign=t]{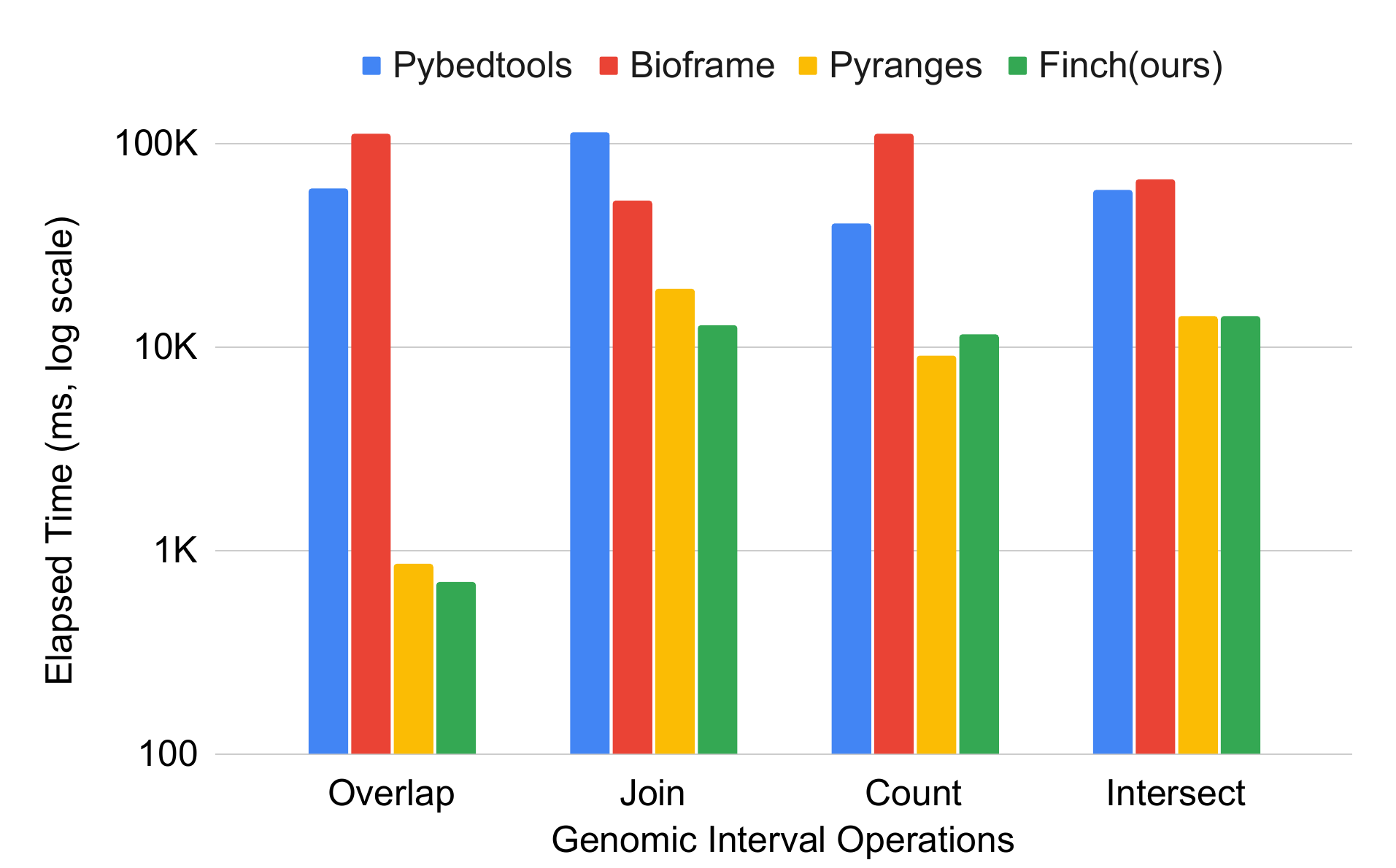}
    \subcaption{Performance comparison using realistic dataset.}
    \label{fig:genomiceval2}
    \end{minipage}
  \vspace{-0.1in}
  \caption{Experimental result of genomic interval operations: lower values indicate better performance.}
  \label{fig:genomiceval}
\end{figure}

Figure~\ref{fig:genomiceval} presents a performance comparison between our generated code using uniform grid and three baseline implementations: Pybedtools~\cite{pybedtools}, Bioframe~\cite{Bioframe}, and Pyranges~\cite{pyranges}. Pybedtools is a Python wrapper of Bedtools~\cite{Bedtools}, a C library which utilizes a hierarchical binning data structure internally, Bioframe leverages the Pandas~\cite{pandas} framework for genomic interval operations, and Pyranges employs a Nested Containment List~\cite{ncls}, a variation of the segment tree written in C. Index building time was not measured in these experiments. 

In Figure~\ref{fig:genomiceval1}, a sensitivity test examines the number of successful intersections using a synthetic dataset. The intervals are uniformly distributed, maintaining a total of 100,000 intervals in both \texttt{Data} and \texttt{Query}. As the x-axis increases, the length of intervals in both \texttt{Data} and \texttt{Query} is extended to increase the number of intersections between intervals. Figure~\ref{fig:genomiceval2} presents a performance comparison on a realistic dataset~\cite{biofast}, with \texttt{Data} containing 8,942,869 intervals and \texttt{Query} containing 1,193,657 intervals. Our generated code demonstrates superior or comparable performance in both synthetic and realistic datasets, with the advantage of being implemented in just 11 lines of code for the Overlap operation, while Bedtools implementation require 206 lines of code.

\subsection{Trilinear Interpolation in Neural Radiance Field}

The third application we explored is a Neural Radiance Field (NeRF)~\cite{nerf} in 3D deep learning. NeRF is a widely used machine learning model in computer graphics and computer vision that generates detailed 3D reconstructions from 2D images. Many NeRF models~\cite{plenoxels,nsvf,nglod} use trilinear interpolation on 3D sparse voxel grids to efficiently represent the 3D scene. In NeRF, rendering a 2D image involves casting rays, sampling points along each ray, performing trilinear interpolation on the sparse voxel grid for each sampled point, and combining these results to compute the final RGB color. Our focus here is on optimizing trilinear interpolation during ray sampling in Plenoxel~\cite{plenoxels}.

\begin{figure}[t]
  \centering
    \begin{minipage}[t]{.6\textwidth}
      \centering
    \raisebox{0.2in}{\includegraphics[width=\textwidth,valign=t]{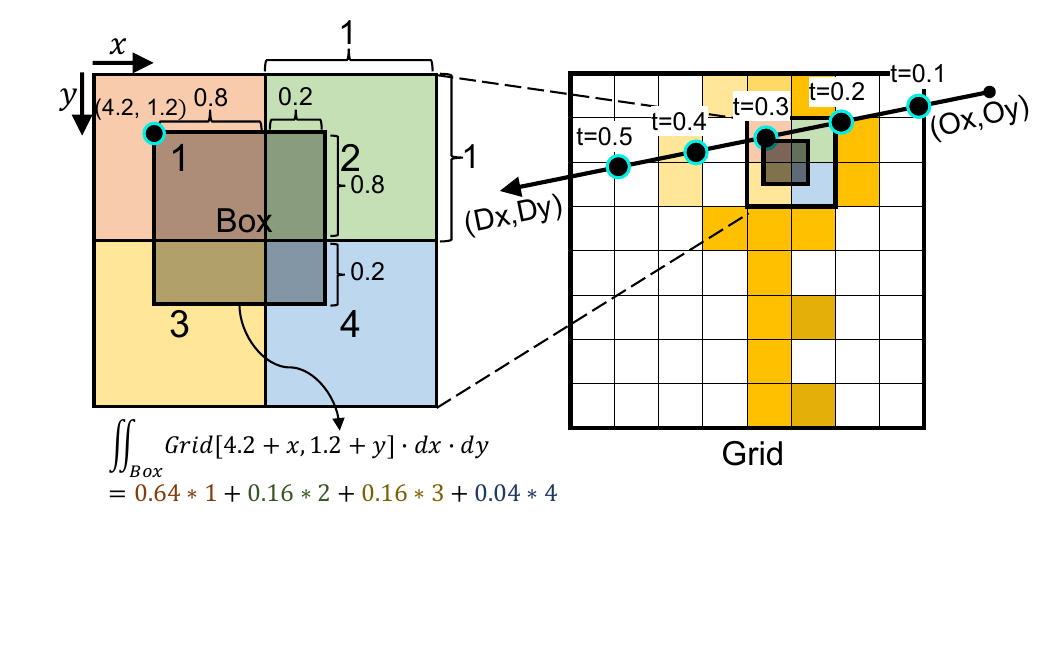}}
    \vspace{-0.55in}
    \subcaption{  \centering Interpolation at sampled points on a sparse grid.\\
    \scalebox{0.9}{$Out_t = Time_t * Grid_{O_x + D_x * t + x, O_y + D_y * t + y} * Box_{x,y} * d(x,y;\mu_\lambda,\mu_\lambda)$}}
    \label{fig:2dnerf}
    \end{minipage}
    \begin{minipage}[t]{.35\textwidth}
    \raisebox{-0.2in}{\includegraphics[width=\textwidth,valign=t]{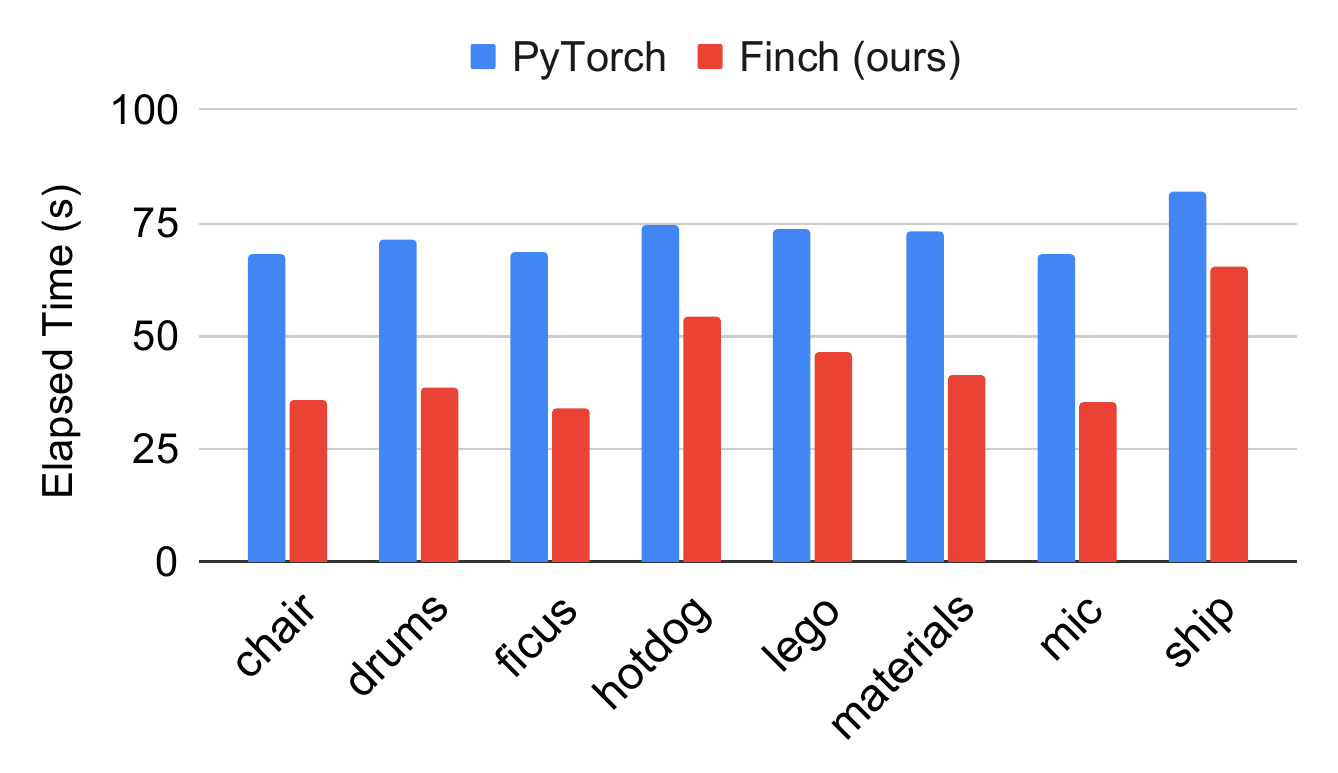}}
    \subcaption{Performance comparison.}
    \label{fig:nerfeval}
    \end{minipage}
  \vspace{-0.1in}
  \caption{(a) Interpolation at sampled points during ray sampling on a sparse grid in 2D for illustrative purposes. (b) Performance comparison between PyTorch and our implementation on various 3D objects in the NeRF Synthetic dataset~\cite{nerf}, where lower values indicate better performance.}
  \label{fig:nerf}
\end{figure}

For explanatory purposes, we illustrate a 2D bilinear interpolation in Figure~\ref{fig:2dnerf}, even though the actual computations are in 3D. This interpolation is applied at each sampled point along the ray by calculating the intersected area using integral reduction:
\[
Out_t = Time_t * Grid_{O_x + D_x * t + x, O_y + D_y * t + y} * Box_{x,y} * d(x,y;\mu_\lambda,\mu_\lambda),
\]
where $(O_x, O_y)$ and $(D_x, D_y)$ are the ray’s origin and direction, respectively. $Box_{x,y}$ is a binary tensor defining the interpolating region, and $Time_t$ is a binary tensor marking the specific time step for sampling along the ray. In practice, these calculations occur in 3D space.

Figure~\ref{fig:nerfeval} presents a performance comparison between our generated code and a PyTorch implementation of trilinear interpolation in Plenoxel. The evaluation was conducted while rendering a 256$\times$256 image using the NeRF Synthetic dataset~\cite{nerf}. Our code achieves a speedup of 1.3$\times$ to 2.0$\times$ over the baseline. This improvement is mainly due to our efficient handling of sparse voxel grids, as we store only non-zero voxels and avoid unnecessary computations. In contrast, the PyTorch implementation uses a fully dense voxel grid, storing even empty voxels and performing calculations regardless of voxel occupancy.

\subsection{3D Point Cloud Convolution}

\begin{figure}
\begin{minipage}[t]{.48\textwidth}
\includegraphics[width=\textwidth,valign=t]{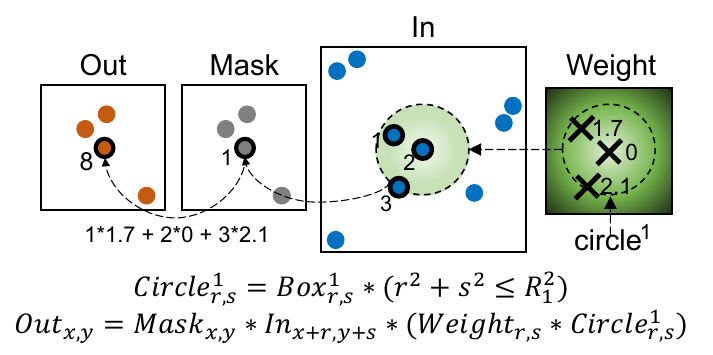}
\subcaption{Kernel Points Convolution (KpConv)~\cite{kpconv}}
\label{fig:conva}
\end{minipage}
\begin{minipage}[t]{.49\textwidth}
\raisebox{-0.08in}{\includegraphics[width=\textwidth,valign=t]{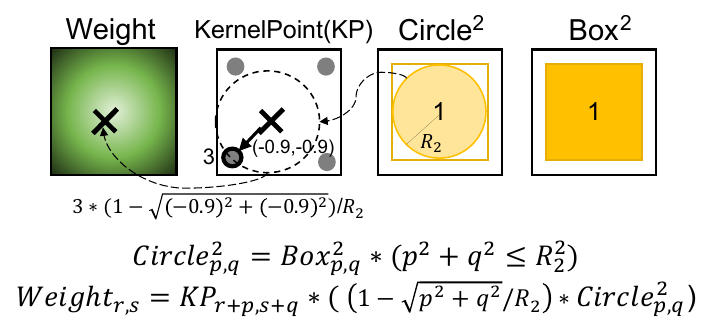}}
\subcaption{Definition of $Weight$. $Weight$ interpolates the points in $KernelPoint$ within radius $R_1$.}
\label{fig:convb}
\end{minipage}
\hspace{0.0\textwidth}
\vspace{-0.15in}
\caption{Kernel Points Convolution on input point clouds written in continuous tensor abstraction. Measure omitted for brevity; all reductions use counting measures.}
\label{fig:kpconv}
\end{figure}

The last application we explored involves 3D point cloud convolution~\cite{tacoucf,minkowskiengine,submanifold}. A 3D point cloud is a set of points represented by $xyz$ coordinates in 3D space, capturing the structure of an object. In 3D deep learning, convolutions are adapted to operate on point clouds rather than grid-like image data, necessitating specialized techniques. KPConv~\cite{kpconv} is a notable example of such a technique, and we demonstrate its implementation using our continuous tensor abstraction.

Figure~\ref{fig:kpconv} shows how KPConv, represented in 2D for simplicity, can be formulated within continuous tensor expression. In Figure~\ref{fig:conva}, KPConv performs convolutions selectively on specific points, marked by a binary mask, $Mask$. For each masked point, the neighboring input points, $In_{x+r,y+s}$, within a radius $R_1$ are gathered and convolved with a continuously varying weight, $Weight_{r,s}$, which depends on the relative position of each neighboring point. The radius-constrained region is indicated by $Circle_{r,s}$, which determines which neighboring points are gathered.

As shown in Figure~\ref{fig:convb}, the continuous weight $Weight_{r,s}$ is defined by interpolating values at fixed set of "kernel points" within a radius $R_2$. This interpolation is distance-based, relying only on kernel points within the radius. Notably, the interpolated weight $Weight_{r,s}$ is not a piecewise-constant, but is evaluated only on pinpoints selected by the mask $Mask_{x,y}$ in the main convolution.

By using continuous tensor abstraction, complex point cloud convolutions that are challenging to implement in traditional tensor programming models can be expressed in a simplified manner. In contrast, the original PyTorch implementation of KPConv requires 2,330 lines of code, including dependencies for neighborhood search libraries~\cite{nanoflann}. With our abstraction, users can specify the whole KPConv operation as a single Einsum:
\small\[
Out_{x,y,m} = Mask_{x,y} * In_{x+r,y+s,c} * (KP_{r+p,s+q,m,c} * (1-\sqrt{p^2 + q^2}/R_{2}^{2}) * Circle^{2}_{p,q}) * Circle^{1}_{r,s} * d(r,s,c;\mu_\#,\mu_\#,\mu_\#)
\] \normalsize

where $m$ and $c$ are indices for the output and input channels, respectively, along with the traversal order of indices and the choice of storage format.

\begin{figure}[t]
  \centering
  \captionsetup{justification=centering}
    \begin{minipage}[t]{.45\textwidth}
    \includegraphics[width=\textwidth,valign=t]{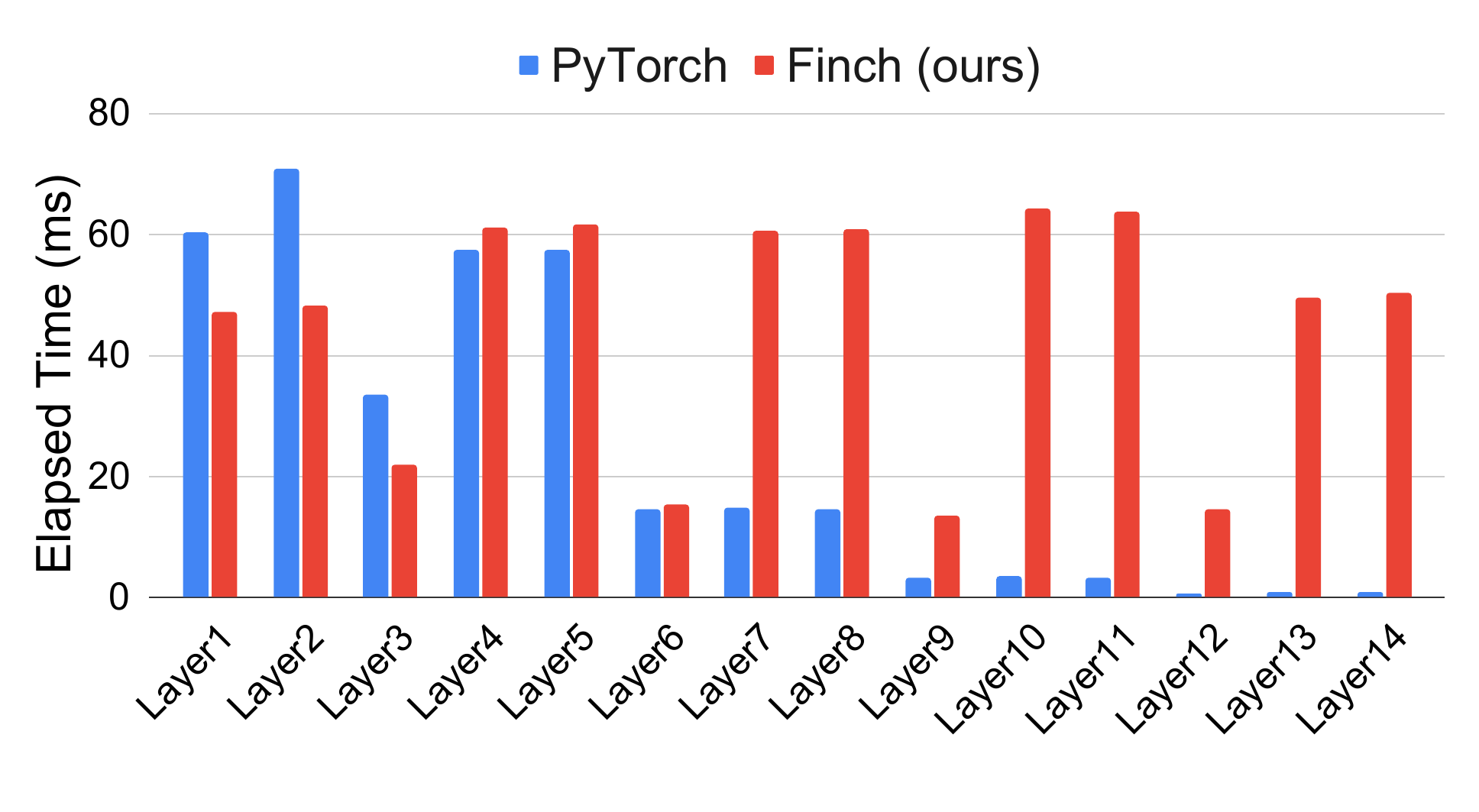}
    \vspace{-0.1in}
    \subcaption{Performance comparison of convolutional layers in the network.}
    \label{fig:pconvevala}
    \end{minipage}
    \hspace{0.1in}
    \begin{minipage}[t]{.45\textwidth}
    \includegraphics[width=\textwidth,valign=t]{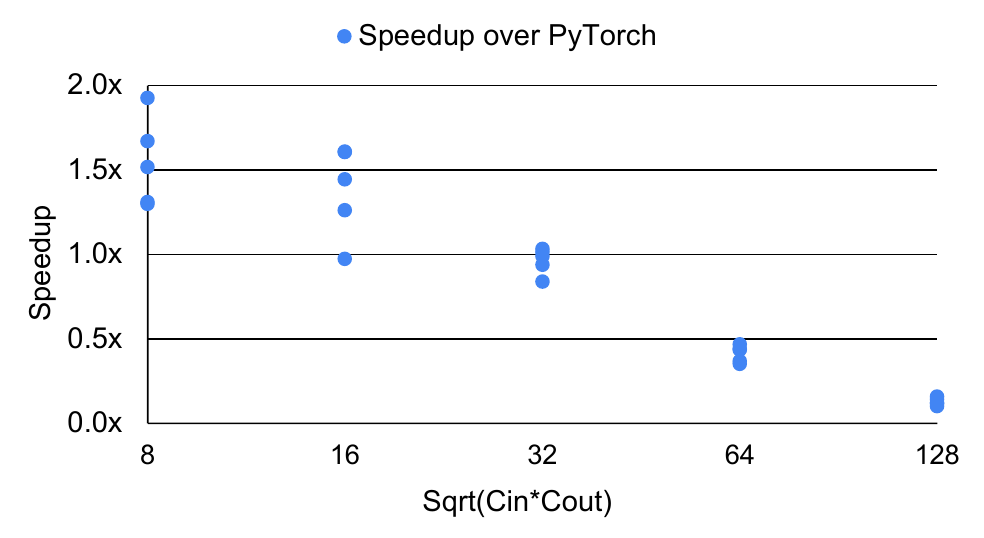}
    \vspace{-0.1in}
    \subcaption{Speedup and slowdown of our generated code over PyTorch with increasing channel size.}
    \label{fig:pconvevalb}
    \end{minipage}
  \vspace{-0.1in}
  \caption{Experimental results of 3D point cloud convolution: (a) Lower is better, (b) Higher is better.}
  \label{fig:pconveval}
\end{figure}

Figure~\ref{fig:pconveval} presents the experimental results comparing our generated code with the PyTorch implementation of KPConv. In Figure~\ref{fig:pconvevala}, the elapsed time of each KPConv layer in the 3D shape classification model architecture using the ModelNet40~\cite{modelnet} dataset is shown. In the initial layers (layers 1-6), our code either outperforms or matches the performance of the PyTorch implementation. However, starting from layer 7, PyTorch begins to outperform our generated code.

We found that the later layers have larger channel sizes, resulting in dense computations. PyTorch leverages highly optimized dense BLAS routines for this large tensor processing. Figure~\ref{fig:pconvevalb} further illustrates this difference by showing the speedup over PyTorch as channel size increases. In the initial layers, where the channel sizes are small ($\sqrt{C_{in} \cdot C_{out}} \leq 32$), our generated code performs better. However, in the later layers with larger channel sizes, where dense computation becomes more significant, PyTorch’s implementation excels. Consequently, our code performs better in the early layers, where the radius query is the primary computational component, while the PyTorch implementation dominates in the later layers, where dense computation is the primary workload.
\section{Related Works}

\para{Dense Tensor Programming Models}
Tensor programming, rooted in Fortran's array data structure~\cite{fortran}, has provided a foundation for diverse applications. In recent years, the machine learning community has introduced frameworks like TensorFlow~\cite{tensorflow}, Jax~\cite{jax}, and PyTorch~\cite{pytorch}, inspired by tensor-focused languages like Matlab~\cite{matlab} and NumPy~\cite{numpy}. These frameworks are instrumental in developing machine learning models, which heavily rely on tensor operations. The latest advancements in scheduling languages~\cite{halide, tvm} have played a pivotal role in enhancing the performance of tensor-based programs. They separate tensor programs into what to compute (algorithm) and how to compute (schedule), simplifying the creation of high-performance tensor programs and the exploration of various loop transformations.

\para{Sparse Tensor Programming Models}
Many of our designs take inspiration from existing sparse tensor programming models. Sparse tensors provide multiple storage formats. The level format abstraction, originally introduced in TACO~\cite{format}, explains the diversity of sparse formats by introducing the concepts of a coordinate hierarchy and level format. This abstraction has further evolved into the fibertree abstraction~\cite{fibertree}, which serves as a format abstraction for our work.

Sparse tensor programming often entails complex code to co-iterate over multiple sparse tensors, each stored in a different format. Numerous compiler projects have dedicated their efforts to generating efficient code for accommodating these diverse formats. Projects like Taichi~\cite{taichi}, MLIR sparse dialect~\cite{mlirdialect}, TACO~\cite{taco}, SparseTIR~\cite{sparsetir}, and Finch~\cite{finch, finchoopsla} can generate efficient code from sparsity-agnostic definitions of computation. The TACO project~\cite{taco,format,sparsearray,tacoschedule} introduces the "merge lattice" concept to efficiently generate code for sparse tensor algebras, even when the tensors are stored in different sparse formats. In a recent development, the Finch project~\cite{finch} introduces the innovative concept of "Looplets," simplifying the generation of sparse code on integer domain through the use of rewriting rules and enhancing extensibility. Looplets support various element types, not limited to numeric types, and a wide range of operators, expanding their versatility. 

\para{Continuous Programming Models}
Several tools, such as Chebfun~\cite{chebfun} and Sympy~\cite{sympy}, offer an intuitive way to manipulate continuous functions in numerical computing. In addition to working with piecewise constant functions, they offer the capability to handle a wider range of function types beyond constant functions. However, their primary focus is not on performance or the tensor programming model. Chebfun focuses on Chebyshev polynomials, a computation class commonly used in numerical computing, which is entirely different from our focus. Both Sympy and Chebfun do not account for sparsity or interval intersections, resulting in the need to compare all pairs of pieces. However, our framework allows the creation of more efficient code that operates only on intersecting pieces, eliminating the need to compare all pairs.

\color{black}

Computer graphics and scientific image visualization languages, such as Diderot~\cite{diderot} and Vivaldi~\cite{vivaldi}, also use continuous fields, enabling them to define functions beyond piecewise-constant, similar to systems like Chebfun or Sympy. However, their primary application often revolves around operations like texture sampling~\cite{pbr}, which fundamentally rely on pointwise evaluations of these continuous fields and their derivatives. Consequently, their design does not prioritize complex geometric operations, and they typically do not support computations on regions beyond pinpoints, such as intervals or N-dimensional shapes. For instance, while Vivaldi supports neighbor queries, these operations are restricted to neighboring points, not continuous intervals. Thus, they lack support for complex geometric operations like interval intersections. In contrast, while our continuous tensor representation is currently restricted to piecewise-constant, our approach enables computations beyond pinpoint evaluation. This allows us to express various geometric operations, such as point cloud convolution and spatial search.

\para{Spatial Database and Spatial Join} 
Spatial databases~\cite{shapely,geos,spatialsql,spatialdatabase} are designed to efficiently handle geometric data. A key operation is the spatial join~\cite{spatialjoin}, which identifies all pairs of geometries from two datasets that satisfy a given spatial predicate (e.g., intersects, contains). This operation is challenging due to the expensive cost of geometric checks and the quadratic number of pairings. Consequently, spatial join has been extensively studied, and various methods have been developed to make it more efficient than exhaustive search. Spatial join is also closely related to partitioning the iteration space in piecewise evaluation semantics. When computing with piecewise tensors, it is essential to efficiently identify pairs of pieces that intersect which is analogous to spatial join. 

A common approach to spatial join is to use spatial indexes, such as interval trees~\cite{intervaltree,ail,ncls,biofast}, KD-trees~\cite{kdtree}, R-trees~\cite{strtree,rtree} or uniform grids~\cite{gridjoin,igd}, to reduce the number of candidate pairs. Tree-based joins traverse hierarchical partition of the space or hierarchical bounding boxes of the geometries~\cite{bvh} to find potentially intersecting pairs. Grid-based joins assign geometries to uniform grid cells, limiting comparisons to objects within the same cell.

Alternatively, some methods avoid spatial indexing by using plane sweep (also known as line sweep) algorithms~\cite{planesweep,planesweep2,spatialjoin}. In this approach, all geometries from both inputs are combined and sorted along a chosen axis. A virtual sweep line then moves across space, maintaining an active set of potentially intersecting intervals. By leveraging this sorted order, plane sweep eliminates the overhead of building spatial indexes and avoids exhaustive pairwise comparisons.

We observe that continuous iteration over real-valued indices in continuous Einsums, together with Looplet-generated code, generalizes the plane-sweep spatial join when co-iterating the piecewise tensors. As shown in the figure, computing $C_x = A_x * B_x$ effectively sweeps a real-valued index $x$ along the $x$-axis of tensors $A$ and $B$.  In higher-dimensional tensors, the storage level order determines the sweeping order of the high-dimensional iteration space.

\begin{figure}[H]
   \begin{subfigure}[t]{0.7\linewidth}
  \includegraphics[width=\linewidth]{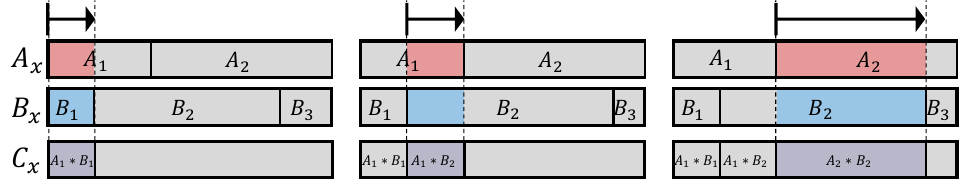}
\caption{$C_x = A_x * B_x$, computed by sweeping a virtual line over the $x$-axis.}
  \label{fig:Looplet}
  \end{subfigure} \hspace{0.1in}
   \begin{subfigure}[t]{0.26\linewidth}
       \vspace{-0.85in} 
\begin{minted}[fontsize=\tiny,escapeinside=||,mathescape=true,frame=single]{python}
pA, pB, pC = 0, 0, 0
while (pA<2 and pB<3):
 lA,rA,vA = A[pA]
 lB,rB,vB = B[pB]
 lAB,rAB = max(lA,lB),min(rA,rB)
 if lAB <= rAB:
  C[pC] = (lAB, rAB, vA*vB)
  pC += 1
 pA,pB += (rA==rAB),(rB==rAB)
\end{minted}
\vspace{-0.05in}
\caption{Generated code }
    \end{subfigure}
    \vspace{-0.1in}
\end{figure}

Unlike the classical plane sweep, which aggregates all geometries, globally sorts them once, and then sweeps—each tensor here keeps its intervals already sorted, and the generated code performs an on-the-fly merge of these sorted streams. Consequently, our sorted storage functions as a lightweight spatial index, and iteration over real indices proceeds in a plane sweep like fashion.

Combining spatial indices with the level format abstraction is a promising direction. Grid-based spatial indices, such as uniform and sparse grids, align naturally with our model. Each grid cell can be seen as an N-D interval, allowing the entire grid to be represented as a continuous tensor. We demonstrated this in our case studies by modeling uniform grids as 2D tensors (Figure~\ref{fig:grid}) and implementing sparse grids in the NeRF example (Figure~\ref{fig:2dnerf}). Tree-based spatial indices like R-trees and KD-trees, are more complex. One way to represent them is by treating the tree as an N-D tensor, where each tree level corresponds to a tensor dimension (e.g., a depth-3 tree as a 3D tensor). Alternatively, the entire tree can represent a single logical dimension by linearizing its nodes (e.g., using pre-order traversal) and wrapping the structure as a single level.

\color{black}
\section{Conclusion}

In this paper, we have introduced the continuous tensor abstraction, which extends tensor indices to real numbers. Our approach, based on piecewise-constant tensors, provides both a novel format abstraction for storage and an efficient code generation technique for continuous tensor expressions. This abstraction enables diverse applications including genomic interval operations, spatial searches, point cloud convolution, and trilinear interpolation on sparse voxel grids. It offers a fresh perspective on these applications, exploring domains largely untouched by traditional tensor programming models. We believe this work opens new  possibilities in tensor programming.

\begin{acks}
We thank reviewers for their valuable feedback.
Supported by Intel; NSF (CCF-2217064, CCF-2107244, CCF-2217099); DARPA (PROWESS HR0011-23-C-0101, SBIR HR001123C0139); DoE PSAAP (DE-NA0003965).
\end{acks}

\bibliographystyle{ACM-Reference-Format}
\bibliography{reference}

\end{document}